\documentclass[reprint,
superscriptaddress,
amsmath,
amssymb,
aps,
nofootinbib
]{revtex4-2}

\usepackage{graphicx,setspace}
\usepackage{dcolumn}
\usepackage{bm}
\usepackage{cancel}
\usepackage{hyperref}
\usepackage{xcolor}
\usepackage{breqn}
\usepackage[normalem]{ulem}
\usepackage{setspace}
\usepackage{ulem}

\usepackage{braket}
\usepackage[titletoc]{appendix}

\usepackage{float}

\renewcommand{\vec}[1]{\mathbf{#1}}

\makeatletter
\let\cat@comma@active\@empty
\makeatother

\begin{document}
\preprint{APS/123-QED}
\title{Tutorial: Shaping the Spatial Correlations of Entangled Photon Pairs}

\author{Patrick Cameron} \email[Corresponding author: ]{p.cameron.1@research.gla.ac.uk}
\affiliation{School of Physics and Astronomy,University of Glasgow, Glasgow G12 8QQ, UK\
}
\author{Baptiste Courme}
\affiliation{Sorbonne Université, CNRS, Institut des NanoSciences de Paris, INSP, F-75005 Paris, France\
}
\affiliation{Laboratoire Kastler Brossel, ENS-Universite PSL, CNRS, Sorbonne Universite, College de France, 24 rue Lhomond, 75005 Paris, France\
}
\author{Daniele Faccio}%
\affiliation{School of Physics and Astronomy,University of Glasgow, Glasgow G12 8QQ, UK\
}%
\author{Hugo Defienne}\email[Corresponding author: ]{hugo.defienne@insp.upmc.fr}
\affiliation{School of Physics and Astronomy,University of Glasgow, Glasgow G12 8QQ, UK\
}%
\affiliation{Sorbonne Université, CNRS, Institut des NanoSciences de Paris, INSP, F-75005 Paris, France\
}


\begin{abstract}

    Quantum imaging enhances imaging systems performance, potentially surpassing fundamental limits such as noise and resolution. However, these schemes have limitations and are still a long way from replacing classical techniques. Therefore, there is a strong focus on improving the practicality of quantum imaging methods, with the goal of finding real-world applications. With this in mind, in this tutorial we describe how the concepts of classical light shaping can be applied to imaging schemes based on entangled photon pairs. We detail two basic experimental configurations in which a spatial light modulator is used to shape the spatial correlations of a photon pair state and highlight the key differences between this and classical shaping. We then showcase two recent examples  that expand on these concepts to perform aberration and scattering correction with photon pairs. We include specific details on the key steps of these experiments, with the goal that this can be used as a guide for building  photon-pair-based imaging and shaping experiments. 
\end{abstract}

\maketitle 

\section{Introduction}

Quantum imaging is the field of optics that aims to exploit the quantum properties of light to improve the performance of imaging systems~\cite{moreau_imaging_2019}. Among the existing non-classical sources of light, those that emit pairs of entangled photons have proven to be particularly fruitful. For example, imaging with sensitivity below the shot-noise limit~\cite{brida_experimental_2010a} and with resolution beating classical diffraction~\cite{boto_quantum_2000,toninelli_resolutionenhanced_2019} were both achieved with photon pair states. In addition, these sources also enabled the development of new imaging modalities. Examples include ghost imaging~\cite{pittman_optical_1995}, imaging with undetected photons~\cite{lemos_quantum_2014}, entanglement-enabled holography~\cite{defienne_polarization_2021} and Hong-Ou-Mandel  microscopy~\cite{ndagano_quantum_2022}. 
Thanks to recent technological advancements, photon-pair-based imaging systems have become relatively simple to implement. Firstly, entangled photon pairs can be produced using spontaneous parametric down conversion (SPDC)~\cite{couteau_spontaneous_2018}. This is a non-linear optical process that converts a single, high-energy photon, called the pump, into two lower energy photons, traditionally called the signal and idler. It does not require complex laser systems or highly specialised non-linear crystals. For example, many experiments — like the one presented here — use standard room-temperature $\beta-$Barium Borate (BBO) crystals and a blue diode laser with a power of a few tens of milliwatts. Then, another factor that adds to the ease of implementation is the recent advent of commercial single-photon-sensitive cameras. Paired with advanced image processing techniques, these cameras enable the detection of photon coincidences across many transverse spatial positions simultaneously — a key measurement when working with entangled photon pairs. For example, electron-multiplying charge-coupled devices (EMCCDs) and single-photon avalanche diodes (SPADs) cameras have enabled the measurement of intensity correlations functions i.e. the $G^{(2)}$ for photon pairs across thousands of spatial modes~\cite{moreau_realization_2012,edgar_imaging_2012,defienne_general_2018,ndagano_imaging_2020}. Despite this recent progress, the low brightness of SPDC sources (i.e. tens of picowatts) and long acquisition times (e.g. sometimes up to several hours) still restrict their application to the laboratory environment and specific niche areas.

Besides generation and detection, it is also important to control how photon pairs and their characteristics (e.g. correlations) propagate within the optical system. For example, spatial light modulators (SLMs) can be placed in the pump beam or the photon pairs path to manipulate their spatial properties. They enable tailoring of the spatial entanglement between photons propagating in free space by manipulating their orbital angular spectrum~\cite{mair_entanglement_2001,walborn_multimode_2003,kovlakov_spatial_2017,kovlakov_quantum_2018,derrico_fullmode_2021,jabir_direct_2017,bornman_optimal_2021}, by shaping momentum-position correlations~\cite{minozzi_biphoton_2013,boucher_engineering_2021,lib_spatially_2020} or by tuning the pump spatial coherence~\cite{zhang_influence_2019,defienne_spatially_2019}. In addition, they can also be used to control photon pair propagation through complex media such as multimode fibers~\cite{defienne_twophoton_2016,leedumrongwatthanakun_programmable_2020,valencia_unscrambling_2020} and scattering layers~\cite{defienne_adaptive_2018,courme_manipulation_2023,devaux_restoring_2022,lib_realtime_2020}. These efforts are mainly directed towards applications in communication and information processing.

Though less explored, such control can also be of interest for quantum imaging. Indeed, if we consider the classical domain, light structuring has many applications in imaging, particularly in microscopy~\cite{maurer_what_2011}. Examples range from structured illumination microscopy~\cite{gustafsson_surpassing_2000}, which aims to improve imaging resolution by illuminating the sample with specifically tailored pattern of light, to adaptive optics~\cite{booth_adaptive_2014} and wavefront shaping~\cite{yoon_deep_2020}, which use an SLM to correct for optical aberrations and scattering, with applications  also for  contrast-enhanced and quantitative phase imaging~\cite{zernike_diffraction_1934,bernet_quantitative_2006}.  
It is therefore interesting to see how we could apply these concepts to quantum imaging and it is within this context that this tutorial is written. Here, we describe two basic experiments that both use an SLM to shape the spatial correlations between entangled photon pairs. We detail their practical implementation and highlight their differences from their classical counterparts. Additionally, we show examples of applications that illustrate the practical uses of this two-photon shaping, including a case in imaging. We also include detailed descriptions of many of the key steps in this work in the Supplementary Material (S.M.), and provide example code in Ref.~\cite{_see_}. 
\begin{figure*}
    \centering
    \includegraphics[width=0.9\textwidth]{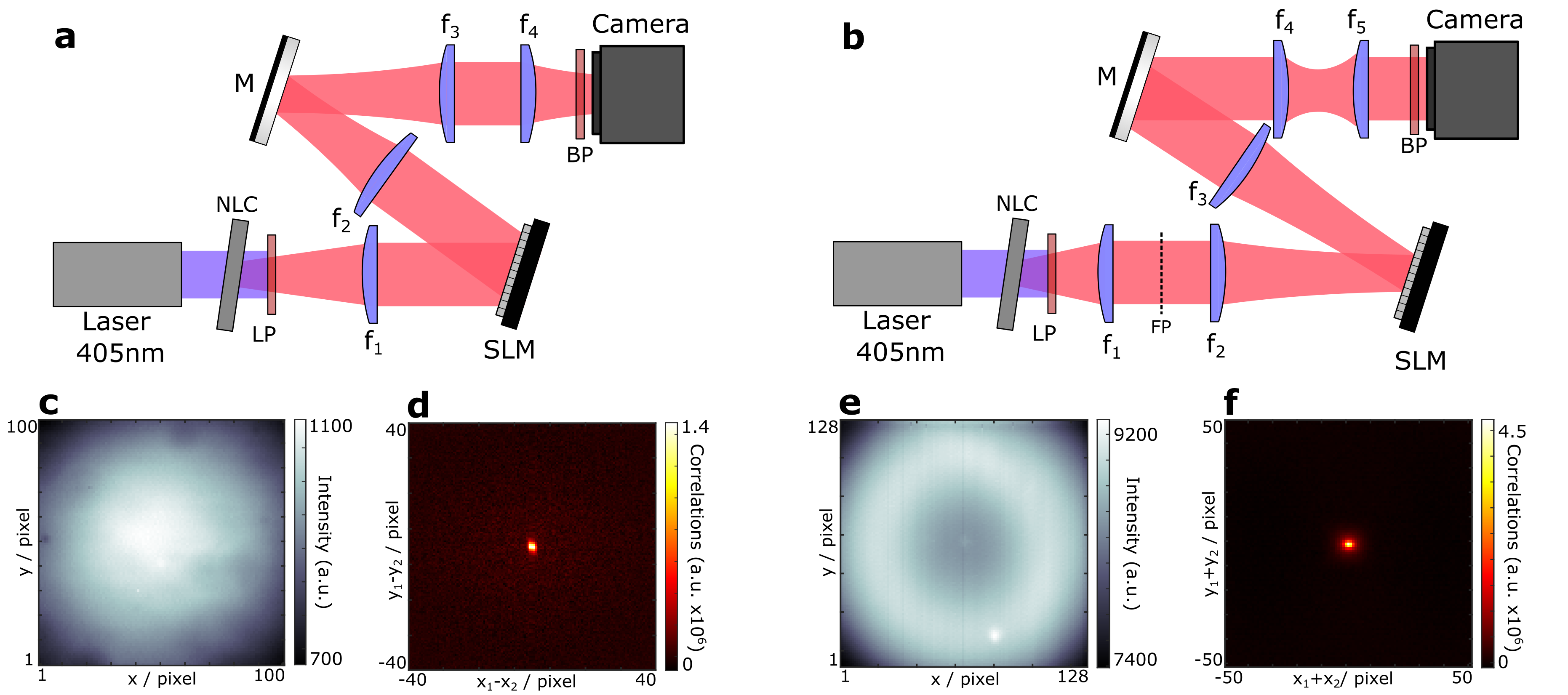}
    \caption{\textbf{Experimental Setup.} Spatially entangled photon pairs centred at $810$nm are produced via Type I spontaneous parametric down conversion (SPDC) using a collimated ($0.8$mm diameter), continuous-wave laser at 405nm and a thin $\beta-$Barium Borate nonlinear crystal (NLC). Pump photons are filtered out by a long-pass filter (LP) at 650nm. A bandpass filter at $810\pm5$nm before the single-photon sensitive camera filters out any non-degenerate pairs. \textbf{a,} Diagram of a far-field shaping configuration (FF) . The surface of the NLC is imaged onto the camera using two 4-f relays ($f_1-f_2$ and $f_3-f_4$). The spatial light modulator (SLM) is placed in the Fourier plane, or far-field, of the crystal. In this configuration, the photons are correlated at the camera, and anti-correlated at the SLM. \textbf{b,} Diagram of a near-field shaping configuration (NF). The Fourier plane (FP) of the NLC, obtained via the lens $f_1$, is imaged into the camera with two 4f relays ($f_2-f_3$ and $f_4-f_5$). The SLM is placed in the conjugate plane to the surface of the NLC. In this configuration, the photons are anti-correlated at the camera, and correlated at the SLM. \textbf{c,e,} Direct intensity images from the camera in the FF shaping (\textbf{c}) and NF shaping (\textbf{e}) configurations. \textbf{d,} Minus-coordinate projection of the measured $G^{(2)}$ in the FF configuration. \textbf{f,} Sum-coordinate projection of the measured $G^{(2)}$ in the NF configuration. \textbf{c,d} are from an acquisition of $2.5\times10^5$ frames. \textbf{e,f} are from an acquisition of $6\times10^6$ frames. M - mirror.}
    \label{fig:experiment}
\end{figure*}

\section{Experimental Setup}

When shaping spatial correlations between pairs of photons, two experimental configurations can be considered. These are called the far-field (FF) and near-field (NF) shaping configurations, named for the position of the SLM relative to the nonlinear crystal. Note that in other works the configurations may instead be named for the position of the detector relative to the crystal. Figures~\ref{fig:experiment}a,b shows schematics of these two configurations. In both cases a nonlinear crystal,  a thin $\beta$-Barium Borate (BBO) crystal, is illuminated by a continuous-wave (CW) pump laser at 405nm. This produces pairs of photons at 810nm via Type I SPDC. Immediately after the crystal, the pump photons are filtered out using a $650$nm-cut-off long-pass filter, allowing only the photons of higher wavelengths generated by SPDC to propagate through the system. In practice, it is common to also include a short-pass filter in the pump beam's path before the crystal (not shown). This filter removes residual low-frequency light emitted by the laser, typically found at the output of blue diode lasers. After propagating through the system, the pairs are detected with an EMCCD camera, and the spatial correlations are measured following the method described in Ref. \cite{defienne_general_2018} (see also Methods). In both configurations, the pairs are controlled by a spatial light modulator that is positioned in a Fourier plane of the camera. A bandpass filter at $810\pm10$nm is used to remove ambient light and most non-degenerate pairs. Due to the phase matching conditions in the SPDC process, the photons at the plane of the crystal are strongly correlated in transverse position ($\vec{r}$) and strongly anti-correlated in transverse momentum ($\vec{k}$)~\cite{walborn_spatial_2010}. Both of these types of correlations can be exploited, hence the two experimental configurations. 

The FF shaping configuration is shown in Figure~\ref{fig:experiment}a. As suggested by the name, the SLM is positioned in a Fourier plane of the crystal, and its surface is imaged on the EMCCD. Since the pairs are correlated in position at the crystal plane, they are also correlated in position at the camera, so they arrive at approximately the same pixel. Figure~\ref{fig:experiment}c shows an example of the intensity at the camera. The shape of this is mostly dependent on the intensity profile of the pump. Figure~\ref{fig:experiment}d shows the so-called minus-coordinate projection of the measured $G^{(2)}$ of the photon pairs. The way this is measured is detailed in the Methods section. It represents the probability of detecting both photons of a pair simultaneously on two pixels of the camera separated by a distance ($x_2-x_1$,$y_2-y_1$), expressed in pixels. This type of measurement is essential when working with entangled photon pairs because it provides information about the spatial correlations between the pairs, something that an intensity image alone does not provide. A bright, narrow peak, as seen in the figure, indicates that the pairs are strongly correlated in position. This means that, when one photon in a pair is detected at a position $(x_1,y_1)$, there is a high probability to detect its twin within a very small area around this pixel $(x_2,y_2)\approx(x_1,y_1)$.  For the measurement of the data shown in Figure~\ref{fig:experiment}d, the SLM did not display any phase mask. 

The NF shaping configuration is shown in Figure~\ref{fig:experiment}b. Here, the SLM is positioned in an conjugate plane to the surface of the crystal, and the Fourier plane of the crystal is imaged onto the camera. Now, due to momentum conservation in the SPDC process, the pairs arrive at the camera at diametrically opposite pixels, relative to the centre of the beam. Figure \ref{fig:experiment}e shows an example of the intensity at the camera. Here, the typical SPDC ring can be seen. The thickness of this ring is proportional to the bandwidth of the pairs, and the radius is dependent on the angle between the pump optical axis and the normal to the crystal surface. Figure~\ref{fig:experiment}f shows the so-called sum-coordinate projection of the measured $G^{(2)}$. The method of measuring this projection is also explained in the Methods section. It represents the probability of detecting two entangled photons at any pair of pixels of the camera $(x_1,y_1)$ and $(x_2,y_2)$ with a given barycenter value $(x_1+x_2,y_1+y_2)$. For example, the central value of the sum-coordinate projection corresponds to the sum of all coincidence rates detected between pairs of pixels with a barycenter at $(0,0)$, which means all the pairs of pixels that are exactly anti-symmetric i.e. $(x_2,y_2)=(-x_1,-y_1)$. In our experiment, the presence of an intense peak at the center indicates strong spatial anti-correlation. This means that when a photon from a pair is detected at a position $(x_1,y_1)$, there is a high probability that its twin will be detected within a very small area around the symmetric pixel $(x_2,y_2) \approx (-x_1, -y_1)$. No phase mask was displayed on the SLM to perform the measurement shown in Figure~\ref{fig:experiment}f. 

\section{Intensity and correlation shaping theory}\label{sec:theory}
In this section we will describe the theory behind two-photon correlation shaping and compare it to classical shaping. When comparing classical and quantum imaging, it is useful to work with the spatial intensity correlation functions, specifically, the first and second order functions $G^{(1)}$ and $G^{(2)}$~\cite{glauber_quantum_1963b}. The quantities we measure in experiment can be written in terms of these, and they can also be used to derive expressions for propagation through an optical system.

First we introduce the relevant measured quantities, and how they are written in terms of $G^{(1)}$ and $G^{(2)}$. When imaging and shaping with classical coherent light, the quantity we measure with the camera is the intensity of the electric field, $I(\vec{r}) = |E(\vec{r})|^2$. This can be written in terms of the first-order spatial correlation function of the field as:
\begin{align} 
    I(\vec{r}) &= G^{(1)}(\vec{r},\vec{r}) \\
    &=  \langle \hat{E}^{(-)} (\vec{r}) \hat{E}^{(+)}(\vec{r}) \rangle,
    \label{eq:intensity_g1}
\end{align}

where $\hat{E}^{(+)}$ and $\hat{E}^{(-)}$ are the positive and negative frequency component of  the quantum operator associated with the electric field, respectively. In practice, the intensity is conventionally measured using a camera by accumulating photons on each pixel.

When working with photon pairs, the quantity that interests us the most is the second-order spatial correlation function of the intensity, $G^{(2)}(\vec{r_1},\vec{r_2})$. This is also often referred as the joint probability distribution of the photons. It is written in terms of the second-order spatial correlation function of the field as:
\begin{equation} 
    G^{(2)}(\vec{r_1},\vec{r_2}) = \langle \hat{E}^{(-)} (\vec{r_1}) \hat{E}^{(-)} (\vec{r_2}) \hat{E}^{(+)}(\vec{r_1}) \hat{E}^{(+)}(\vec{r_2}) \rangle.
    \label{eq:intensity_g2}
\end{equation}

The $G^{(1)}$ describes the field correlations and the second-order correlation function can be seen as the intensity-correlation analogue of this. In practice, it is measured by detecting photon coincidence between pairs of spatial positions $\vec{r_1}$ and $\vec{r_2}$. As detailed in Ref.~\cite{defienne_general_2018}, it can also be reconstructed by measuring the intensity covariance between pairs of pixels on a camera. Assuming we have pure two-photon states, $G^{(2)}(\vec{r_1},\vec{r_2})$ gives the probability of simultaneously detecting two photons from a pair at positions $\vec{r_1}$ and $\vec{r_2}$. This quantity is related to the spatial two-photon wave function $\phi$, as: $G^{(2)}(\vec{r_1},\vec{r_2}) = |\phi(\vec{r_1},\vec{r_2})|^2$. In this way, $\phi$ is to $G^{(2)}$ in the two-photon case as the field $E$ is to $I$ in the case of coherent light.

Since we want to see how we can shape the distributions $I$ and $G^{(2)}$, it is useful to know how they propagate. Any linear system can be described by its coherent point-spread function (PSF). Given the PSF, written $h(\vec{r}',\vec{r})$, the correlation functions can be propagated through any linear system as: 
\begin{align}\label{eq:prop_g1full}
    G^{(1)}(\vec{r_1}',\vec{r_2}')= \int G^{(1)}&(\vec{r_1},\vec{r_2})h^*(\vec{r_1}',\vec{r_1}) \nonumber \\
    &h(\vec{r_2}',\vec{r_2}) d\vec{r_1}d\vec{r_2}
\end{align}
and
\begin{align}\label{eq:prop_g2full}
    G^{(2)}&(\vec{r_1}',\vec{r_2}',\vec{r_3}',\vec{r_4}') = \int G^{(2)}(\vec{r_1},\vec{r_2},\vec{r_3},\vec{r_4})h^*(\vec{r_1}',\vec{r_1}) \nonumber \\ 
    & \times h^*(\vec{r_2}',\vec{r_2}) h(\vec{r_3}',\vec{r_3})h(\vec{r_4}',\vec{r_4})d\vec{r_1}d\vec{r_2}d\vec{r_3}d\vec{r_4}.
\end{align}

For example, let's consider the two experimental configurations described in Figures~\ref{fig:experiment}a and b (NF and FF shaping). In both cases, the camera is positioned in a Fourier plane of the SLM. Since the action of the SLM is to impart a phase profile, $\theta(\vec{r})$, onto the beam, the PSF associated with the propagation between the SLM (shaping plane) and the camera (detection plane) can be written as
\begin{equation}\label{eq:psf_shaping}
    h(\vec{r}',\vec{r}) = \mathrm{exp}\left[-\frac{2\pi i \vec{r}\vec{r'}}{f\lambda} + i\theta(\vec{r}) \right],
\end{equation}
where $f$ is the focal length of the lens immediately after the SLM, $\lambda$ is the wavelength of the light being used, $\vec{r}$ and $\vec{r}'$ are the transverse coordinates in the SLM and camera plane, respectively. Note that in each configuration there are in reality several lenses performing a magnification in addition to the Fourier transform, which can be taken into account by changing the effective value of $f$ in the previous equation.

\begin{figure*}
    \centering
    \includegraphics[width=0.6\textwidth]{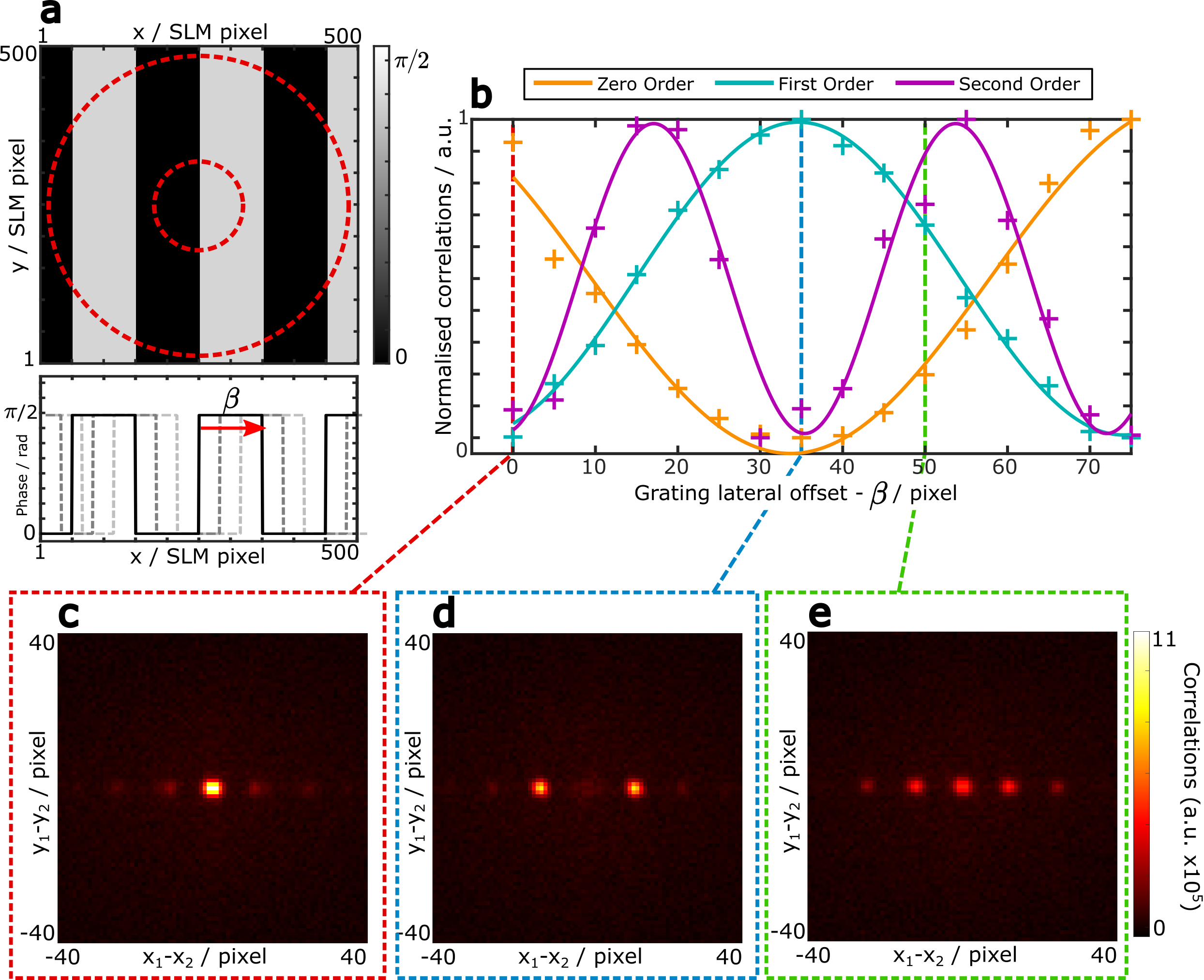}
    \caption{\textbf{Shaping the two-photon correlations in a far-field shaping (FF) configuration.} A phase grating (\textbf{a}) is displayed on the illuminated region of the SLM. Line plot shows a single row of the grating. Red dashed circles are illustration of approximate position of SPDC ring on the SLM. Grating width is exaggerated for illustrative purposes. The actual period used in acquisition was 75 pixels. \textbf{b,} Magnitude of correlation peaks as a function of grating lateral offset $\beta$. Crosses are data points, solid lines are sine fits. Data is normalised for each order individually to allow better visual comparison. \textbf{c-e,} Minus-coordinate projections of $G^{(2)}$ for $\beta=0$ (\textbf{c}), $\beta=35$ pixels (\textbf{d}), and $\beta=70$ pixels (\textbf{e}) peaks (cropped). Correlations for each grating offset are from an acquisition of $\sim1.8\times10^6$ frames.}
    \label{fig:nf imaging}
\end{figure*}

\subsection{Shaping coherent light}

For a perfectly spatially coherent source, e.g. a laser, the first-order correlation function is given by 
\begin{equation} \label{eq:g1_coherent}
    G^{(1)}_{coh}(\vec{r_1},\vec{r_2}) = E^*(\vec{r_1})E(\vec{r_2}),
\end{equation}
where $E(\vec{r})$ is the electric field at position $\vec{r}$ in the SLM plane and $^*$ denotes the complex conjugate. Thus, from equations \ref{eq:intensity_g1} and \ref{eq:prop_g1full}, the intensity after propagation through the system is given by 
\begin{equation}
    I_{coh}(\vec{r}') = \left| \int E(\vec{r})h(\vec{r}',\vec{r})d\vec{r} \right|^2.
\end{equation}
Using equation \ref{eq:psf_shaping} and simplifying, we obtain the intensity at the camera
\begin{equation}\label{eq:shapingCoherent}
     I_{coh}(\vec{u}) = \left |\mathcal{F} \left[E(\vec{r})\right]*\mathcal{F}\left[e^{i\theta(\vec{r})}\right] \right|^2\left(\frac{\vec{r}'}{f\lambda}\right),
\end{equation}
where $\mathcal{F}[...]$ is the 2-dimensional Fourier transform, and $*$ denotes the 2-dimensional convolution operation. This is the expected result from Fourier-optics and says that the field measured at the camera is simply the (scaled) Fourier transform of the phase mask on the SLM. The intensity correlation function for a coherent source is simply
\begin{equation}
    G^{(2)}_{coh}(\vec{r_1},\vec{r_2}) = I_{coh}(\vec{r_1})I_{coh}(\vec{r_2}).
\end{equation}
For coherent light, $G^{(2)}$ can thus be fully computed from the intensity measurements and therefore does not contain any additional information.

\subsection{Shaping entangled photon pairs}
Now let's consider a two-photon state. In the input plane i.e. the SLM plane, we can write our state in the position basis as 
\begin{equation}\label{eq:state}
    \ket{\phi} = \iint \phi(\vec{r_1},\vec{r_2})\ket{\vec{r_1},\vec{r_2}} d\vec{r_1} d\vec{r_2},
\end{equation}

where $\phi(\vec{r_1},\vec{r_2})$ is the two-photon wavefunction expressed in the position basis, $\vec{r_1}$ and $\vec{r_2}$ are the transverse positions in the SLM plane for photon 1 and photon 2 respectively. In our case, both photon have the same polarisation and spectral frequency. This results from the pair generation process that we have chosen to use in our experiments, which is Type I SPDC. In this simplified notation, $\ket{\vec{r_1},\vec{r_2}}=\ket{\vec{r_1}}_1\otimes\ket{\vec{r_2}}_2$ denotes the state in which photon 1 is at position $\vec{r_1}$ and photon 2 is at position $\vec{r_2}$. The $G^{(1)}$ for such a state is given by 
\begin{equation}
    G^{(1)}_{pairs}(\vec{r_1},\vec{r_2})=\int \phi^*(\vec{r_1},\vec{r})\phi(\vec{r_2},\vec{r}) d\vec{r}.
\end{equation}
and the direct intensity is then 
\begin{equation} 
    I_{pairs}(\vec{r_1}) = \int \left|\phi(\vec{r_1},\vec{r}) \right|^2d\vec{r}.
\end{equation}
Additionally, one can compute the second-order field correlations as:
\begin{equation}
    G^{(2)}_{pairs}(\vec{r_1},\vec{r_2},\vec{r_3},\vec{r_4}) = \phi^*(\vec{r_1},\vec{r_2})\phi(\vec{r_3},\vec{r_4}),
\end{equation}
giving the intensity correlations:
\begin{equation}
    G^{(2)}_{pairs}(\vec{r_1},\vec{r_2}) = \left|\phi(\vec{r_1},\vec{r_2})\right|^2.
\end{equation}

Using equations \ref{eq:prop_g1full} and \ref{eq:prop_g2full}, $I_{pairs}$ and $G^{(2)}_{pairs}$ can be expressed in the camera plane:
\begin{equation}  \label{eq:shapingPhotonPairs1}
   I_{pairs}(\vec{r_1}') = \int \left|\int \phi(\vec{r_1},\vec{r_2}) h(\vec{r_1}',\vec{r_1}) d\vec{r_1}\right|^2 d\vec{r_2},
\end{equation}
and
\begin{equation}  \label{eq:shapingPhotonPairs2}
    G^{(2)}_{pairs}(\vec{r_1}',\vec{r_2}') = \left | \int \phi(\vec{r_1},\vec{r_2})h(\vec{r_1}',\vec{r_1})  h(\vec{r_2}',\vec{r_2}) d\vec{r_1} d\vec{r_2} \right|^2,
\end{equation}
where $\vec{r_1}'$ and $\vec{r_2}'$ are the transverse positions in the camera plane, $\phi(\vec{r_1},\vec{r_2})$ is the two-photon wavefunction in the SLM plane and $h$ is the PSF.

Now we consider the two shaping configurations shown in Figures~\ref{fig:experiment}a and b. In both cases, the imaging system from the SLM to the camera, $h$, is the same (up to a different magnification factor), but the input states are different. We start with the configuration of Figure~\ref{fig:experiment}a. Here, the SLM is positioned in the Fourier plane of the crystal. In our experimental conditions the collimated pump beam diameter is much larger than the crystal thickness. Therefore, we can assume that photon pairs are near-perfectly anti-correlated in the plane immediately before the SLM~\cite{abouraddy_entangledphoton_2002}. That is, the wavefunction $\phi(\vec{r_1},\vec{r_2}) \approx \phi_0(\vec{r_1}-\vec{r_2})\delta(\vec{r_1}+\vec{r_2})$, where $\phi_0$ is the amplitude envelope of the two-photon wavefunction in the crystal Fourier plane. It is linked to the intensity measured in the SLM plane as $I(\vec{r}) = |\phi_0(\vec{r})|^2$. In practice, it takes the shape of a disk or a ring, as shown in Figure~\ref{fig:experiment}.e, and its spatial phase is assumed to be uniform. After performing the change of variables $\vec{r_+} = (\vec{r_1}+\vec{r_2})/2$ and $\vec{r_-} = (\vec{r_1}-\vec{r_2})/2$, the intensity correlation in the camera plane can be expressed as
\begin{equation} \label{eq:2photonShaping1}
    G^{(2)}_{pairs}(\vec{r_+'},\vec{r_-'})= \left| \mathcal{F}\left[\phi_0(\vec{r})\right]*\mathcal{F}\left[e^{i\psi(\vec{k})}\right] \right|^2 \left(\frac{2 \vec{r_-'}}{f \lambda} \right)
\end{equation}
where $\psi(\vec{k})=\theta(\vec{k})+\theta(-\vec{k})$ and all global constants are omitted for clarity.

Now we consider the configuration in Figure \ref{fig:experiment}b. Here we are imaging the crystal plane onto the SLM. Since we have thin crystal and a large pump diameter, we can assume that the photon pairs are perfectly correlated, i.e. $\phi(\vec{r_1},\vec{r_2}) \approx \phi_0'(\vec{r_1}+\vec{r_2})\delta(\vec{r_1}-\vec{r_2})$~\cite{abouraddy_entangledphoton_2002}, where $\phi_0'$ is the amplitude envelope of the two-photon wavefunction in the crystal plane. It is linked to the intensity measured in the SLM plane as: $I(\vec{r}) = |\phi_0'(\vec{r})|^2$. In practice, it takes the shape of the pump beam, as shown in Figure~\ref{fig:experiment}.c, and its spatial phase is assumed to be uniform. Then, the intensity correlations at the camera are given by
\begin{equation} 
\label{eq:2photonShaping2}
     G^{(2)}_{pairs}(\vec{r_+'},\vec{r_-'})= \left| \mathcal{F}\left[\phi_0'(\vec{r})\right]*\mathcal{F}\left[e^{i 2\theta(\vec{k})}\right] \right|^2 \left(\frac{2 \vec{r_+'}}{f \lambda} \right).
\end{equation}
By performing a similar propagation calculation with $G^{(1)}_{pairs}$, we find that the intensities on the camera plane are spatially uniform in both configurations and do not depend on the phase displayed on the SLM i.e. $I_{pairs}(\vec{r'})$ is a constant. Thus, as would be the case for a perfectly spatially incoherent source, the SLM does not modulate the intensity measured in the Fourier plane. This near-perfect spatial incoherence is indeed observed experimentally and is a consequence of our experimental conditions, specifically the use of a collimated pump with a diameter much larger than the thickness of the crystal. By changing the crystal and the illumination conditions~\cite{saleh_duality_2000a}, it would be possible to work in an intermediate regime with partially spatially coherent light, allowing modulation of both the intensity and the intensity correlations.

Comparing equations \ref{eq:shapingCoherent}, \ref{eq:2photonShaping1} and \ref{eq:2photonShaping2} we see that the spatial intensity correlations can be shaped in a manner that is almost equivalent to spatial intensity shaping in the classical case. Under our experimental conditions, $\mathcal{F}[E(\vec{r})]$, $\mathcal{F}[\phi_0(\vec{r})]$ and $\mathcal{F}[\phi_0'(\vec{r})]$ are very sharply peaked, and can therefore be approximated as Dirac Delta functions. This allows for the simplification of the equations to focus on the role played by the SLM:
\begin{itemize}
    \item  Classical: $I_{coh}=\left|\mathcal{F}\left[e^{i\theta(\vec{r})}\right]\right|^2$, \
    \item  Pairs - FF shaping: $G^{(2)}_{pairs} = \left|\mathcal{F}\left[e^{i\psi(\vec{r})}\right]\right|^2$,\
    \item Pairs - NF shaping: ${G^{(2)}_{pairs}}=\left|\mathcal{F}\left[e^{i2\theta(\vec{r})}\right]\right|^2.$
\end{itemize}

Interestingly, with entangled photon pairs we find that depending on the configuration, the shaping behaves slightly differently. In the case of FF shaping, we see that the phase on the SLM affects the projection in the form of the function $\psi(\vec{r})=\theta(\vec{r})+\theta(\vec{-r})$. This suggests that what shapes the spatial correlations is effectively the SLM mask plus a spatially inverted version of the mask. In the case of NF shaping, we see that the projection is affected by a $2\theta(\vec{r})$ phase term, i.e. the pairs `see' a phase that is double that of the actual mask on the SLM. 
\begin{figure*}
    \centering
    \includegraphics[width=0.6\textwidth]{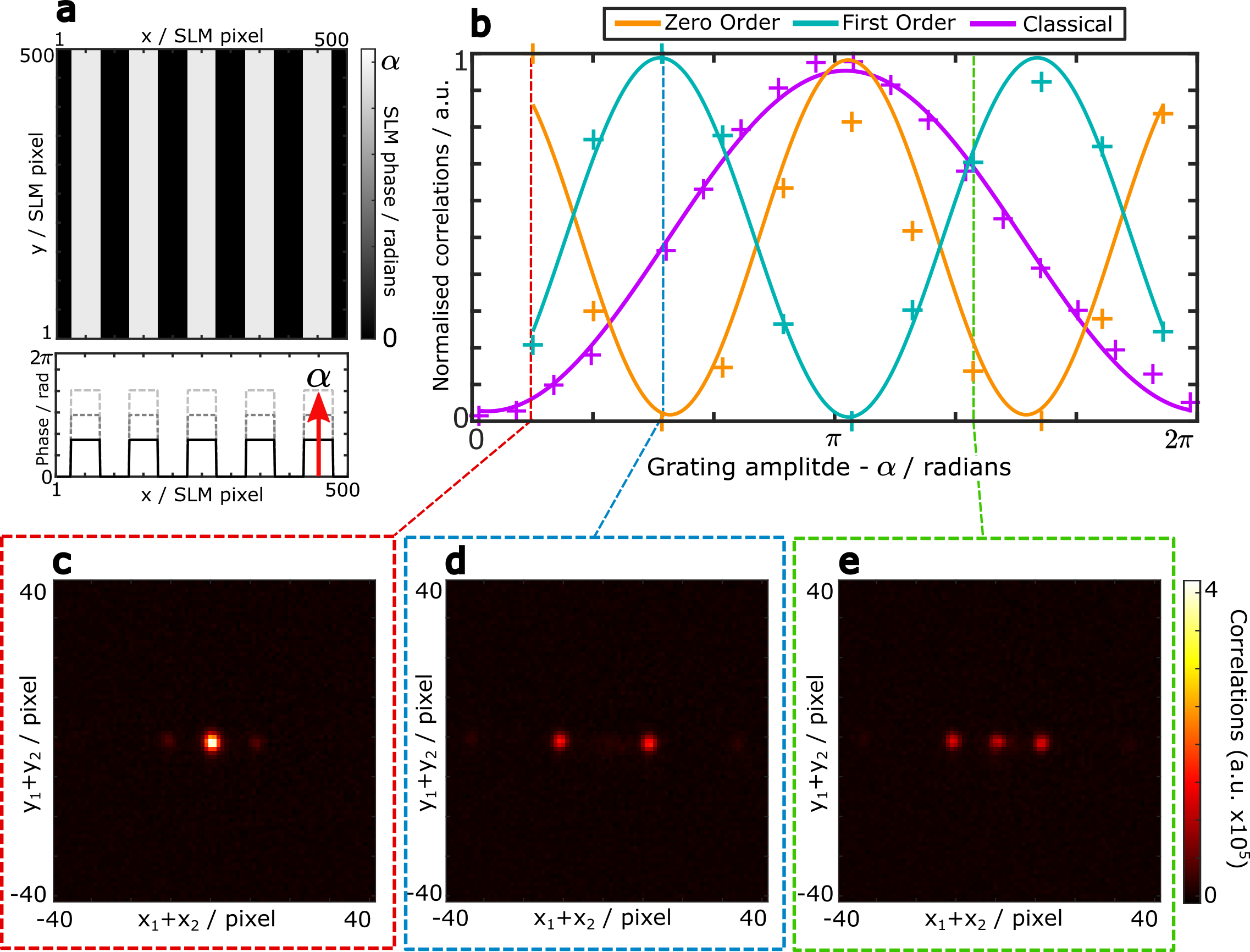}
    \caption{\textbf{Shaping the two-photon correlations in a near-field shaping (NF) configuration. a,} Phase grating displayed on the illuminated region of the SLM. Line plot shows a single row of the grating. \textbf{b,} Magnitude of correlation peaks as a function of grating amplitude $\alpha$. Crosses are data points, solid lines are sinusoidal fits. Data is normalised for each order individually to allow better visual comparison. Classical data (purple) has scale of intensity, rather than correlations. \textbf{c,d} Sum-coordinate projections of $G^{(2)}$ for grating amplitude  $\alpha=\pi/6$ (\textbf{c}), $\alpha=\pi/2$ (\textbf{d}) and $\alpha=2\pi/9$ (\textbf{e}). Correlations for each grating amplitude are from an acquisition of $\sim5\times10^5$ frames. For details on the acquisition of the classical data, see S.M.}
    \label{fig:ff imaging}
\end{figure*}

\section{Results}
We now show how to experimentally measure and modulate the photon-pair $G^{(2)}$, following equations~\ref{eq:2photonShaping1} and \ref{eq:2photonShaping2}. For this, we perform shaping experiments using the FF and NF configurations shown in Figures~\ref{fig:experiment}a and b, respectively. 

From a practical standpoint, it is important to first clarify a key aspect regarding the measurement of the spatial $G^{(2)}$ that we perform. The $G^{(2)}$ is measured using the approach described in Ref.~\cite{defienne_general_2018} and in Methods. This quantity takes the form of a four-dimensional matrix. To visualize it, we project it into two-dimensional images as seen in examples in Figures~\ref{fig:experiment}d and f. Depending on whether we use the FF or NF configuration, we perform projections {onto} the {minus-}~or sum-coordinates, respectively. These projections are formally defined as spatial averages of the measured $G^{(2)}$ function: $C^-(\vec{r}_-)= \int_S G^{(2)}(\vec{r_-},\vec{r_+}) d\vec{r_-}$
for the minus-coordinate projection, and $C^+(\vec{r}_+)= \int_S G^{(2)}(\vec{r_-},\vec{r_+}) d\vec{r_-}$ for the sum-coordinate projection, where $G^{(2)}$ is expressed here using the variables $(\vec{r_+},\vec{r_-})$ and $S$ is the integration area. As seen in equations \ref{eq:2photonShaping1} and \ref{eq:2photonShaping2}, these projections follow the symmetries of the measured $G^{(2)}$ that we expect in each configuration:
\begin{align}
    C^-(\vec{r}_-) &\propto \left|\mathcal{F}\left[ e^{i\psi(\vec{r})}\right]\right|^2 \quad\mbox{(FF)}\\
    C^+(\vec{r}_+) &\propto  \left|\mathcal{F}\left[e^{i 2 \theta(\vec{r})}\right]\right|^2 \quad\mbox{(NF)} 
\end{align}
These projections effectively reveal the spatial modulation of $G^{(2)}$, which is what we aim to observe, while maximising the signal-to-noise ratio in the measured signal. In practice, this spatial averaging reduces the acquisition time needed to resolve the correlations from several hours to just a few minutes.

Figure~\ref{fig:nf imaging} shows the results for the FF shaping configuration, where we are imaging the near-field of the crystal on the camera (Fig. \ref{fig:experiment}a). As a simple demonstration, we use a $\pi/2$-modulated phase grating pattern, shown in Figure~\ref{fig:nf imaging}a. According to equation \ref{eq:2photonShaping1}, the photon pairs in this system `see' the actual grating plus a spatially inverted version. Therefore, if $\theta(\vec{r})+\theta(\vec{-r})=const$, then we expect to see no modulation of the minus-coordinate projection. To demonstrate this, the grating is displayed on the SLM and translated laterally, measuring $C^-$ at each lateral shift. Figure~\ref{fig:nf imaging}b shows the values of each diffraction order as a function of lateral shift. If the grating is positioned such that one of the steps is at the centre of the SPDC beam, then we have $\psi(\vec{r}) =\pi/2=constant$ and we expect no modulation in the ideal case. Figure~\ref{fig:nf imaging}c shows the minus-coordinate projection measured in such a case. Even though we do not observe a complete extinction of higher orders, mainly due to the fact that the phase pattern is not perfectly asymmetric, in practice the measurements in Figure~\ref{fig:nf imaging}a confirm that these higher orders are minimized while the zero order is maximized. Instead, if the grating is positioned $1/4$ of a grating period away from this, then we get $\psi(\vec{r})=2\theta(\vec{r})$, and the correlations should be maximised in the first-order diffraction peaks. This can be seen in Figure~\ref{fig:nf imaging}d. At the intermediate grating positions, $\psi(\vec{r})$ contains higher frequency components, and we observe second-order diffraction peaks. Furthermore, it's interesting to note that replicating this experiment with classical coherent light would not yield any change in the intensity-measured diffraction pattern. Indeed, a lateral shift in the phase pattern in the Fourier plane would only influence the spatial phase component of the diffraction pattern, which the camera is not sensitive to.

According to equation~\ref{eq:2photonShaping2}, we expect a different phenomenon in the case of the NF shaping configuration (Fig.~\ref{fig:experiment}b), {where we are now imaging the Fourier plane of the crystal on the camera}. As before, we display a phase grating on the SLM (Fig~\ref{fig:ff imaging}a) but, since we instead have the $2\theta(\vec{r})$ phase term, lateral translation will have no effect on $C^+$. Instead, we vary the amplitude of the phase grating $\alpha$ from 0 to $2\pi$ radians, and measure $C^+$ at each step. Figure \ref{fig:ff imaging}b shows the values of the diffraction orders at each grating amplitude. As a comparison, we also performed the same experiment using classical coherent light and recorded the first-order diffraction peak intensity in function of $\alpha$ (pink curve). As expected, it follows a sinusoidal pattern reaching its maximum at $\alpha = \pi$. In the case of entangled photon pairs, we see the oscillation but, since $C^+(\delta\vec{r_+})$ is modulated by twice the phase mask, the frequency of this oscillation is doubled. Figures~\ref{fig:ff imaging}c-e show examples of the sum-coordinate projections measured for $\alpha$ equals to {$\pi/6$, $\pi/2$ and $2\pi/9$}, respectively. 

Finally, it is important to note that in se two experiments described, the intensity images measured on the camera do not change regardless of the phase pattern displayed on the SLM: they remain the same as that shown in Figures~\ref{fig:experiment}c and e.

In light of these results, it is interesting to discuss the physical interpretation of such a $G^{(2)}$ shaping demonstration, in particular using the corpuscular picture. By modulating $G^{(2)}$, we effectively change the joint probability of detecting both photons of a pair at two specific camera pixels. However, we achieve this without changing the probability of detecting a photon at a given camera pixel (i.e. the marginal probability), since the intensity remains uniform. For instance, in the case of Figure~\ref{fig:nf imaging}d, the probability has been modified in such a way that the photons no longer arrive on the camera strongly correlated in position; instead, they are now forced to be separated from each other by a distance defined by the period of the phase grating programmed on the SLM. In essence, this means that we can control the collective behaviour of photon pairs without changing their individual behaviour.

\begin{figure*}
    \centering
\includegraphics[width=0.9\textwidth]{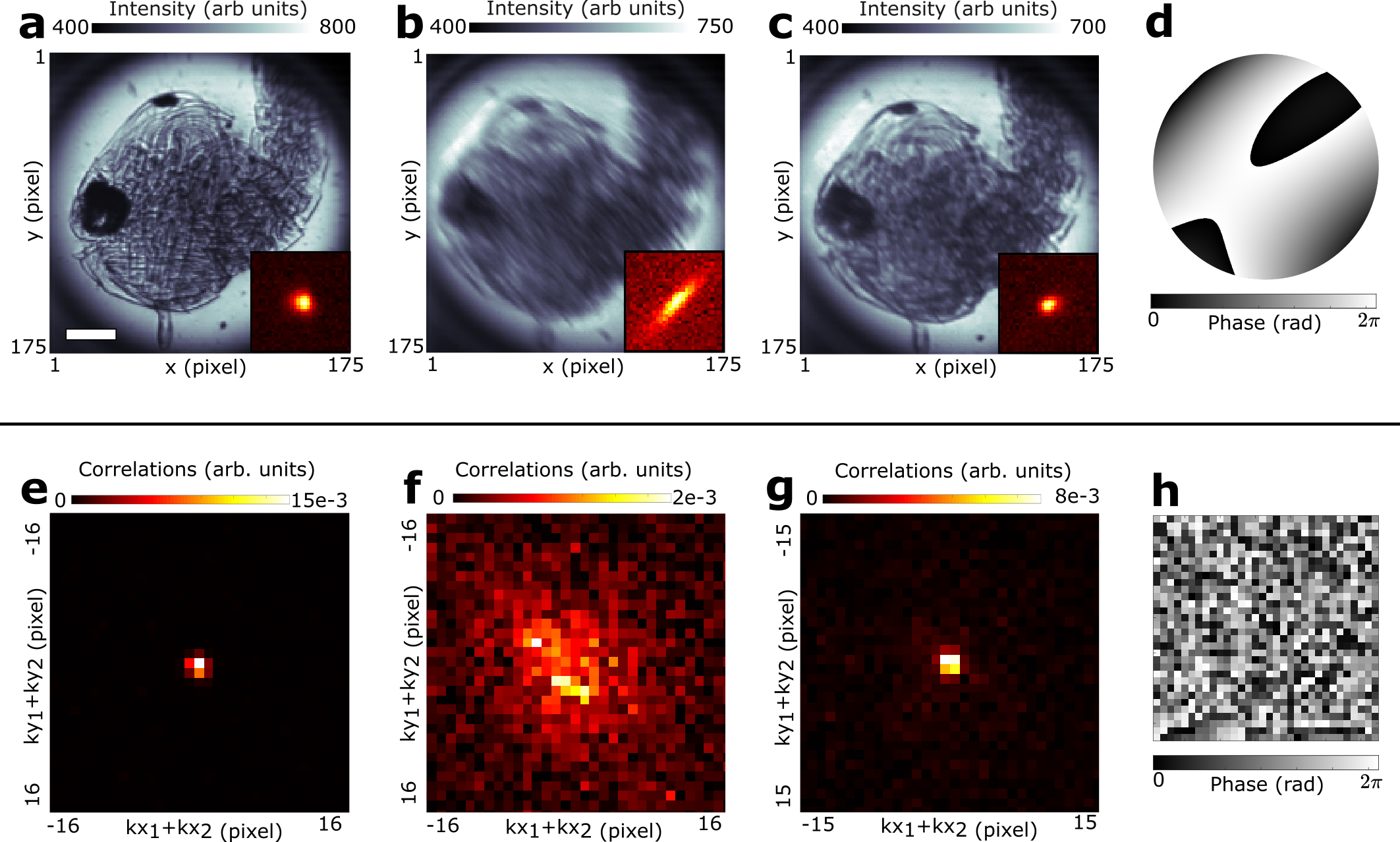}
    \caption{\textbf{Examples of applications of two-photon shaping.}\textbf{a-d} Example 1: Adaptive optics with entangled photons. The sample, a mosquito pupa, is illuminated with photon pairs and imaged onto an EMCCD camera. The direct intensity (large grayscale images) and sum-coordinate projection (inset images) are measured. A layer of transparent polymer (polydimethylsiloxane) placed after the sample is used to introduce aberrations, which distorts the intensity images and projection. An optimisation algorithm is used to find the phase mask on an SLM that maximises the projection central value, which also restores imaging performance. The SLM is positioned in the conjugate image plane of the crystal (i.e. NF shaping configuration). A detailed experimental setup can be found in Ref.~\cite{cameron_quantumassisted_2023}. \textbf{a,} Reference intensity and sum-coordinate projection of $G^{(2)}$ images with no aberrations present. \textbf{b,} Intensity and projection images in the presence of aberrations. \textbf{c,} Intensity and projection images after the projection peak has been optimised. \textbf{d} Example phase mask retreived by optimising the projection peak. \textbf{e-h}, Example 2: Entanglement transmission through a scattering medium. A thin scattering medium is placed in the Fourier-plane of a BBO crystal. The transmission matrix of this medium is measured using a classical beam, and is used to compute a correction mask displayed on an SLM in a conjugate plane to the scatterer and the crystal surface (i.e. NF shaping configuration). Without the scatterer, the spatial entanglement of the pairs can be verified. With the scatterer, the entanglement can no longer verified. Applying the correction on the SLM allows measuring entanglement again after the scattering medium. \textbf{d}, Sum-coordinate projection without the scatter. \textbf{e}, Sum-coordinate projection of $G^{(2)}$ in the presence of the scatterer. \textbf{f}, Sum-coordinate projection with the scatterer and after correction. \textbf{h,} Example phase mask retreived by inverting the classically measured transmission matrix.  A detailed experimental setup can be found in Ref.~\cite{courme_manipulation_2023} }
    \label{fig:applications}
\end{figure*}
\section{Applications}

In this section we show two examples of experiments that use these concepts of two-photon shaping. The first is the method of quantum-assisted adaptive optics (QAO), as reported in Ref.~\cite{cameron_quantumassisted_2023}. In this work the spatial correlations of entangled photons are exploited to gain information about the optical aberrations within an imaging system. Effectively, the central value of the sum-coordinate projection is used as a guidestar to implement an adaptive optics (AO) scheme. 
The key result underpinning this is effectively an extension of Equation \ref{eq:2photonShaping2}. In the low-aberration (i.e. non-scattering) regime, the sum-coordinate projection directly encodes the point-spread function of the system via 
\begin{equation}
    C^+(\delta\vec{r_+})\propto|h*h|^2,
\end{equation}
where $*$ denotes the convolution operation. This means that if we can find the phase mask on an SLM that optimises $C^+$, it will also optimise $h$ and restore imaging performance. Figure \ref{fig:applications}a-c shows results of an AO experiment using this QAO method. The experimental setup used to obtain these results is the same as that described in Ref.~\cite{cameron_quantumassisted_2023}, where the SLM is located in an image plane of the crystal, and the camera is positioned in a Fourier plane (i.e. NF shaping configuration). This approach is interesting because, unlike many classical AO methods, it does not rely on choosing a good image metric and the performance is independent of the sample structure. Additionally, it should be noted that the quantum correlations only need to be used as an optimisation target and the actual imaging can be fully classical, avoiding the long acquisition times typically involved in quantum imaging experiments. 

In some imaging systems, aberrations become so complex that we enter in the regime of light scattering. Correcting optical scattering using classical light is a complex task~\cite{vellekoop_focusing_2007,popoff_measuring_2010} and it becomes even more challenging when using photon pairs. Indeed, each photon of the pairs is scattered in random directions, causing them to lose their correlations. Since these correlations are at the core of photon pair-based quantum imaging, preserving them is essential for these systems to function. This situation is explored in Ref.~\cite{courme_manipulation_2023}, which we show here as an example. In this work, the authors propagate a two-photon state - identical to that generated in the experiments shown in Figures~\ref{fig:experiment}a and b - through a scattering medium (parafilm layer). The SLM is located in an image plane of the crystal, and the measurement is performed in a Fourier plane (i.e. NF shaping configuration). The experimental setup is described in full detail in Ref.~\cite{courme_manipulation_2023}. As we can see in the sum-coordinate projections shown in Figures~\ref{fig:applications}d.e., the strong spatial correlations between the pairs measured in free-space (Fig.~\ref{fig:applications}d) are lost in the presence of the scattering medium (Fig.~\ref{fig:applications}e).

To restore the correlations, the authors leverage the concepts of transmission matrix based wavefront shaping, introduced in Ref.~\cite{popoff_measuring_2010}. The transmission matrix, $T$, is the discrete version of the point-spread function. In this formalism the two-photon state $\Psi^{in}$ is propagated {with matrix multiplications. For a system with an SLM followed by the scattering medium, we have:}
\begin{equation}
    \Psi^{out} = TD\Psi^{in}D^tT^t,
\end{equation}
where $D$ is a diagonal matrix representing propagation through the SLM and $\Psi^{out}$ is the matrix associated with the two-photon state at the output. In this method, the transmission matrix is first measured classically using a coherent light source at 810nm, the same wavelength as the photon pairs. Then, once the transmission matrix is known, it can be used to compute a phase mask {for} the SLM that negates the effects of the scattering layer. This, in turn, restores the spatial correlations of the pairs. Figures \ref{fig:applications}e show the sum-coordinate projection measured after applying the correction mask on the SLM. 

Both of these methods aim to recover spatial correlations by correcting for optical aberrations. In the first example, the correction is found by directly optimising the quantum correlations. This is possible because it targets the case when the aberrations are low-order (i.e. smooth), meaning the correlation signal ($C^+$) can be resolved in a relatively short time ($\sim2$ minutes). If the aberrations become sufficiently high-order so that there is scattering, as is the case with the second example, then the acquisition times become prohibitive. Therefore, the correction is found via a classical transmission matrix measurement instead. 

\section*{Conclusion}
In summary, we have shown how the spatial correlations of an entangled two-photon state can be shaped with both a theoretical description and an experimental examples. Structuring such correlations is very similar to shaping intensity with classical light, but with some important differences. Depending on the position of the SLM relative to the crystal (i.e. the spatial basis in which we modulate the phase of the photons, position or momentum), the photon pairs `see' a different version of the phase mask displayed on the SLM. For an SLM {located} in the position basis (i.e. conjugate plane to the crystal surface), correlations are modulated according to twice the phase mask. For an SLM positioned in the momentum basis (Fourier plane of the crystal), they are modulated by the combination of the phase mask and its spatial inverse. We demonstrate this experimentally, showing that there are in fact cases where we see no modulation of the correlations even when a non-flat phase mask is displayed on the SLM. 

The aim of this article is to introduce spatial correlation-shaping between entangled photons with a simple example that has a well-known classical analogue. We provides the necessary technical details for readers to  replicate the described experiments. We believe that these simple experiments can be expanded to inspire new methods of quantum imaging, much like what has been done in the examples shown in Figure~\ref{fig:applications}. 

\section*{Acknowledgements}
\noindent \textbf{Funding:} D.F. acknowledges support from the Royal Academy
of Engineering Chairs in Emerging Technologies Scheme
and funding from the United Kingdom Engineering
and Physical Sciences Research Council (Grants No.
EP/M01326X/1 and No. EP/R030081/1) and from the
European Union Horizon 2020 research and innovation program under Grant Agreement No. 801060. H.D. acknowledges funding from the ERC Starting Grant (Grant No. SQIMIC-101039375). P.C and H.D. acknowledge support from SPIE Early Career Researcher Accelerator fund in Quantum Photonics. \\
\section*{Author Contributions}
P.C, B.C, and H.D performed the experiments and analysed the results. P.C and H.D designed the experiments. H.D and D.F supervised the project. All authors discussed the data and contributed to the manuscript. 

\section*{Methods}

\subsection{Details on the experimental setup}
The experimental setup is depicted in Figure~\ref{fig:experiment}. The pump is a collimated continuous-wave laser at 405nm (Coherent OBIS-LX) with an output power of 200mW and a beam diameter of 0.8$\pm0.1$ mm. The BBO crystal has dimensions $0.5\times5\times5$ mm and is cut for Type I SPDC at 405nm with a half opening angle of 3 degrees (Newlight Photonics). The crystal is slightly rotated around the horizontal axis to ensure near-colinear phase matching of photons at the output. A 650 nm cut-off long pass filter is used to remove the pump photons after the crystal. This is important, as the pump can cause fluorescence in the subsequent lenses. A band-pass filter at $810\pm10$nm selects near-degenerate photon pairs. The SLM is a PLUTO-NIR-15 from Holoeye. It is a liquid-crystal-on-silicon device with a resolution of 1920$\times$1080 pixels, and a pixel pitch of $8\mu$m. The beam radius on the SLM is approximately 2.4mm, corresponding to 300 pixels. The EMCCD is the model iXon Ultra 897 from Andor, which has a resolution of 512x512 pixels with a pixel pitch of 16$\mu$m. The camera is operated with a region of interest (ROI) of $100\times100$ pixels. 

\noindent \textbf{Far-field shaping configuration}. In Figure~\ref{fig:experiment}a, for clarity, lens $f_1$ is depicted as a single lens; however, in reality it represents three lenses arranged in the confocal configuration 50 mm - 150 mm - 100 mm. The first lens is positioned 50 mm after the crystal, and the last lens is situated 100 mm before the SLM. Lenses $f_2-f_4$ each have a focal length of 100 mm, with $f_2$ placed 100 mm after the SLM and $f_4$ placed 100 mm before the camera. 

\noindent \textbf{Near-field shaping configuration}. { In Figure~\ref{fig:experiment}b, for clarity, lenses $f_1$ and $f_2$ are depicted as two lenses; however, in reality it represents four lenses arranged in the confocal configuration 45 mm - 75 mm - 50 mm - 150 mm.  The first lens is positioned 45 mm after the crystal, and the last lens is situated 150mm before the SLM. Lenses $f_3$, $f_4$ and $f_5$ have a focal length of 100 mm, 50 mm and 75 mm, respectively, with $f_3$ placed 100mm after the SLM and $f_5$ placed 75 mm before the camera. }

\subsection{Measurement $G^{(2)}$, $C^-$ and $C^+$}\label{app:measuremnets}

In each experiment of our work, we measure the spatially-resolved second-order intensity correlation function $G^{(2)}$, and then use it to compute the sum {and minus} coordinate projections. $G^{(2)}$ takes the form of a 4-dimensional matrix containing $(N_x\times N_y)^2$ pixels, where $N_x \times N_y$ corresponds to the region of the sensor used to capture data. An element of the matrix is written $G^{(2)}_{ijkl}$, where $(i,j)$ and $(k,l)$ are pixel labels corresponding to spatial positions $(x_i,y_j)$ and $(x_k,y_l)$. {Assuming that the two-photon state detected by the camera is pure, $G^{(2)}_{ijkl}$ can be obtained directly by measuring the covariance matrix, as detailed in Ref.~\cite{defienne_general_2018}}.{Such a matrix} is measured by acquiring a set of $M+1$ frames, denoted $\{I^{(l)}\}_{l\in[\![1,M+1]\!]}$, using a fixed exposure time of $0.002$s and then processing them using the formula: 
\begin{equation}\label{eq:g2computation}
    G^{(2)}_{ijkl} = \frac{1}{M}\sum^M_{l=1}\left[ I^{(l)}_{ij}I^{(l)}_{kl} - I^{(l)}_{ij}I^{(l+1)}_{kl}\right].
\end{equation}
The first term in the sum is the covariance matrix between a frame and itself and corresponds to an estimate of the real coincidences from photon pairs. The second term is the covariance matrix between a frame and a subsequent frame. Since we know the true pairs will never be detected across two different frames, this corresponds to an estimate of accidental coincidences coming from uncorrelated photons being detected in the same frame.

Since the EMCCD camera cannot resolve the number of photons incident on a single pixel, the photon coincidences at the same pixel cannot be measured, and so the corresponding values $G^{(2)}_{ijij}$ are set to the mean of the neighbouring values. $G^{(2)}_{ijkl}$ is a discrete version of the continuous second-order intensity correlation function $G^{(2)}(\mathbf{r_1},\mathbf{r_2}) = |\phi(\mathbf{r_1},\mathbf{r_2})|^2$ where $\phi$ is the spatial two-photon wave-function associated with the photon pairs. Such a formalism has been employed in many studies describing the propagation of entangled photon pairs \cite{fedorov_gaussian_2009, jamesschneeloch_introduction_2016,abouraddy_entangledphoton_2002}. 
Then, all the coordinate projections can be computed from $G^{(2)}$. $C^-(\vec{r_-})$ is defined as 
\begin{equation}
    C^-(\vec{r_-}) = \int G^{(2)}_{\pm}(\vec{r_+},\vec{r_-}) d\vec{r_+}
\end{equation}
where $G^{(2)}_{\pm}$ is $G^{(2)}$ expressed using the variables $(\vec{r_+},\vec{r_-})$, such that $G^{(2)}_{\pm}(\vec{r_+},\vec{r_-}) = G^{(2)}(\vec{r_1},\vec{r_2})$. Using the variables $(\vec{r_1},\vec{r_2})$, the equation becomes: 
\begin{equation}
    {C^-(\vec{r_-}) = \int G^{(2)}(\vec{r_1},\vec{r_1}-2 \vec{r_-}) d\vec{r_1}.}
\end{equation}
{Similarly, $C^+(\vec{r_+})$ is defined as:}
\begin{align}
   { C^+(\vec{r_+})} &= {\int G^{(2)}_{\pm}(\vec{r_1}, \vec{r_-}) d\vec{r_1}} \\
    &= {\int G^{(2)}(\vec{r},2 \vec{r_+}-\vec{r_1}) d\vec{r_1}}.
\end{align}
In the manuscript, we use indifferently $G^{(2)}_{\pm}$ and $G^{(2)}$ for clarity. In practice, $C^-$ is calculated using the discrete-variable formula: 
\begin{equation}
    C^-_{i^-j^-} = \sum^{N_x}_{i=1}\sum^{N_y}_{j=1} G^{(2)}_{(i-i^-)(j-j^-)ij},
\end{equation}
and $C^+$ using:
\begin{equation}
    C^+_{i^+j^+} = \sum^{N_x}_{i=1}\sum^{N_y}_{j=1} G^{(2)}_{(i^+-i)(j^+-j)ij}.
\end{equation}

\subsection{Example of $G^{(2)}$ Measurement in Practice}

The previous section gives a mathematical description of a $G^{(2)}$ measurement, but it is also useful to have a more practical description. 

For simplicity, let us assume that we have $N_x=N_y=N$ so each frame contains $N^2$ pixels. A typical acquisition requires around 10 million frames. It is not practical - and normally not even possible - to store this many frames for each acquisition, so instead we break the processing down into smaller blocks and continuously sum the results of each block. The real and accidental coincidence terms from Equation~\ref{eq:g2computation} are computed separately and the subtraction is done after the acquisition. The $m^{th}$ block results in two $N^2\times N^2$ arrays corresponding to these terms, called $R^{(m)}$ and $A^{(m)}$ respectively. These are summed for each block to get the final arrays so that $R=\sum_mR^{(m)}$ and $A=\sum_mA^{(m)}$, and finally $G^{(2)}=R-A$. 

The following is a step by step breakdown of the acquisition procedure:
\begin{enumerate}
    \item Initialise empty $N^2\times N^2$ arrays corresponding to $R$ and $A$. $R^{(m)}$ and $A^{(m)}$ will be added to these arrays after each block, then discarded. 
    \item Fill the camera's internal buffer with frames and download them as a block to the processing computer. \
    \item When the buffer has been emptied begin the next acquisition to fill the buffer. At the same time, begin the processing on the most recent block of frames. See following for details on processing.\ 
    \item Add $R^{(m)}$ and $A^{(m)}$ to $R$ and $A$ and delete the current block of frames from memory. \ 
    \item Return to Step 2 and repeat until the required number of frames have been processed.\ 
    \item Finally, compute $G^{(2)}=R-A$ and, if required, compute $C^{+}$ or $C^{-}$ from $G^{(2)}$.
\end{enumerate}

The processing for each block is done as follows. For the $m^{th}$ block of $M$ frames, labelled $I^{(m)}$, will be in the form of an $N^2\times M$ array where each column is a frame that has been unwrapped into a 1-dimensional vector. From this, $R^{(m)}$ can be computed via an outer product. In practise, this is simply a matrix multiplication with the array of frames: 
\begin{equation}
    R^{(m)}=\frac{1}{M}I^{(m)}{I^{(m)}}^T,
\end{equation} 
where $X^T$ denotes the transpose of $X$.

$A^{(m)}$ is computed slightly differently. We define $I^{(m)}_{1}$ as $I^{(m)}$ with the last frame (i.e. column) removed. Similarly, we define $I^{(m)}_{2}$ as $I^{(m)}$ with the first frame (column) removed. $I^{(m)}_1$ and $I^{(m)}_2$ are both $N^2\times (M-1)$ arrays. Now, the accidental coincidence matrix is computed as
\begin{align}
    A^{(m)}=\frac{1}{2(M-1)}\left( I^{(m)}_1{I^{(m)}_2}^T+I^{(m)}_2{I^{(m)}_1}^T\right).
\end{align}
The two terms are the covariance matrix of a given frame with the next frame, and a given frame with the previous frame. 
For $N\times N$ pixel frames, $G^{(2)}$ is actually an $N\times N \times N \times N$ element 4-d array. However, this processing gives $G^{(2)}$ in an $N^2\times N^2$ form, where each column/row corresponds to a 2-d conditional probability distribution that has been unwrapped into a 1-d column/row vector. 

This processing is done in parallel to the acquisition of the next block of frames. For an EMCCD camera, the processing is faster than the acquisition, so the camera speed is still the limiting factor. See Ref.~\cite{_see_} for a template MATLAB script to perform the acquisition and processing. At the end of the processing we typically save $R$, $A$, and the relevant projection. The intensity images taken from the camera are all discarded as they are processed. Therefore, we also typically save a total intensity image that is simply the sum of all of the frames that have been acquired.
\subsection{Alignment}

Aligning a photon pair-based imaging experiment can be difficult since the power of the down-converted light is on the order of {tens of picowatts}, making it invisible to the eye, and most cameras. In addition to this, it is essential to optimise the correlation width of the pairs, otherwise signal-to-noise will be prohibitively low. This is very sensitive to lens misalignment, meaning rougher alignment methods that may work in classical imaging systems such as simply positioning lenses with a ruler, will not be sufficient.

In this section we will briefly describe the alignment process. Due to the low flux, it is impractical to align the system directly with the SPDC light. Instead, it is easier to use a bright, spatially coherent source at the same wavelength as the pairs, e.g. a laser. Here, we use a superluminescent diode (SLED) that is spatially coherent, and has a spectrum of $\sim800-820$ nm. The same bandpass filter as the one used to filter photon pairs is used to select the desired wavelength at $810$ nm from this range. 

If this alignment beam is co-aligned with the SPDC pump laser, then co-propagating photon pairs will follow the same general trajectory, and so it can be used to align the optics after the nonlinear crystal. The lens directly after the crystal should be relatively short - i.e. have a sufficiently high NA - to ensure all of the high-wavevector photons are collected. The position of this lens is critical, so mounting it on a translation stage for fine control is recommended. The position of the crystal itself is also key, so it is recommended that the crystal also be mounted on a stage. In general, the lenses should be placed starting from the camera, and working backwards, finishing with the short lens after the crystal. For a full description of the alignment process, see SM.

\subsection{Calibrating the SLM}

An SLM is an array of liquid crystal pixels whose birefringence is controlled by an applied voltage. Effectively, it is a display on which grayscale images can be show. The grayscale value at each pixel determines the voltage applied across the pixel, and therefore controls its birefringence. This variation in birefringence across the SLM imparts a spatially dependent phase to light that is incident on it. Ideally, the imparted phase will be a linear function of pixel value, with grayscale value of 0 giving 0 phase shift, and 255 giving 2$\pi$. In practise, this is not generally the case. The imparted phase is wavelength-dependent, so a grayscale range of 0-255 could correspond to a phase range of more or less than 0-2$\pi$. The response may also be non-linear. To account for nonlinearity, we calibrate the pixel response via the decorrelation of a reference speckle as a function of pixel grayscale value. {Note that there are plenty of other methods to calibrate an SLM, such as the more common one described in Ref.~\cite{balondrade_calibration_2018}. The one we describe here has the advantage of being less sensitive to alignment errors.}

Two things are needed for this method: a spatially coherent source of illumination at the desired wavelength (e.g. a SLED), and a scattering medium. The coherent light must be (roughly) collimated at the SLM, and illuminate a sufficiently large region. The scattering medium can be anything that produces a speckle pattern. A thin piece of semi-transparent plastic e.g. from a plastic sleeve, or a layer of parafilm stretched across a microcope slide works well. By placing the scatterer in the beam path, we {measure} a reference speckle at the camera. Then, by randomly selecting approximately half of the SLM pixels and scanning their grayscale value from 0 to 255, the resulting speckle images will be changed relative to the reference as a function of the actual phase imparted. {One can demonstrate that the value of spatial correlation ($corr2$ function of Matlab) between this speckle and the reference speckle varies like the cosine of the global phase between the two speckles.} With this information, a calibration curve can be found {to determine the correct relationship between SLM gray scale values and optical phase values}. For a full description of the calibration process, see SM. Example code and data can be found in~\cite{_see_}. 

\bibliographystyle{apsrev4-2}
\bibliography{references}

\newpage
\onecolumngrid
\appendixpage
\appendix

\hspace{5pt}
\begin{figure*}[ht]
    \centering
    \includegraphics[width=0.9\textwidth]{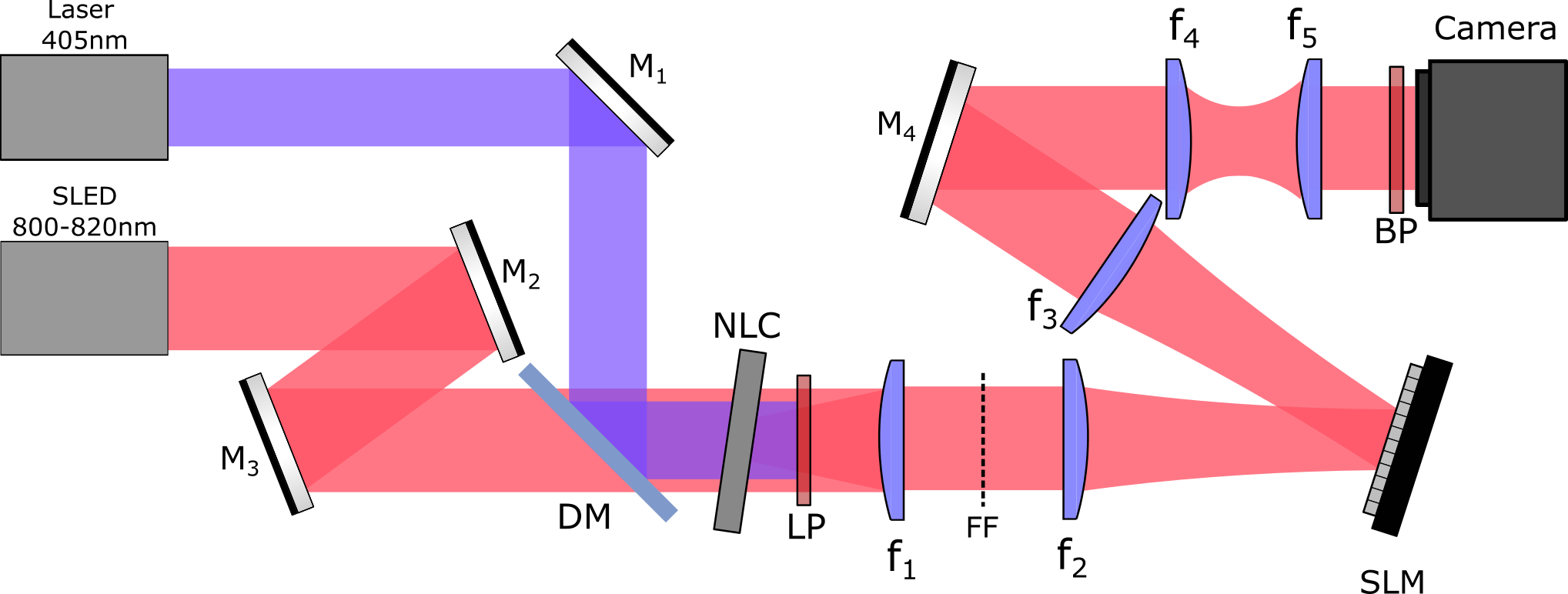}
    \caption{\textbf{Experimental setup including classical alignment beam.} A superluminescent diode (SLED) is superimposed onto the path of the pump laser using a dichroic mirror (DM). Mirror $M_1$ and the DM provide full position and angle control of the pump. Mirrors $M_2$ and $M_3$ provide the same for the SLED. In this configuration, the SLED is collimated at the SLM, enabling the demonstration of classical coherent shaping. }
    \label{fig:experimentSM}
\end{figure*}

\section{Details on classical shaping experiment}

The classical shaping data shown in Figure \ref{fig:ff imaging}c was acquired with the experiment in the configuration shown in Figure \ref{fig:experimentSM} with the 405 nm pump laser turned off. Here, for clarity, the $4$-$f$ imaging system $f_1-f_2$ is represented by two lenses, but in reality is composed of 4 lenses in a confocal arrangement with focal lengths 50 mm - 150 mm - 100 mm - 200 mm. The first and last lenses are positioned at focal distances away from the crystal and SLM, respectively. The lenses following the SLM are also in a confocal arrangement, with focal lengths $f_3=150$ mm, $f_4=50$ mm, $f_5=150$ mm. Lens $f_3$ is placed 150mm after the SLM, and lens $f_5$ is placed 150mm before the camera, in this case a standard CCD (charge-coupled device). 

\section{Additional information on the {correlation} measurement {with an EMCCD camera}}

In this section we include any relevant extra information regarding the $G^{(2)}$ measurement:

\begin{itemize}
    \item $G^{(2)}$ will be an array of $(N_x\times N_y)^2$ elements, where $N_x,N_y$ are the dimensions of the acquired images. For most computers, $N_x,N_y\gtrsim 200$ will result in a JPD that is too large to be stored in RAM. Therefore, it is useful to use a region-of-interest (ROI) on the camera to reduce the images to a manageable size if necessary.
    \item For correlation measurements, we operate the EMCCD camera at temperatures around $-60^oC$. However, this temperature setting may be unstable if the framerate is too high. If the sensor temperature is unstable, it will increase the noise and affect the pixel response, which in turn will hinder the correlation estimation. It is therefore recommended to ensure that the temperature is stable by reducing the framerate. This can be done by either increasing the integration time of each frame or adding a pause between frames to allow the sensor to cool. 
    \item The pixels in camera sensors typically suffer from crosstalk, i.e. the readout value of one pixel can be affected by neighbouring pixels. This results in artificial correlations between neighbouring pixels that is usually much greater that the real photon pair correlations. For EMCCDs, there is a charge-smearing effect in the readout process. This causes pixels in the same row to be strongly correlated. Therefore, the correlation values between these pixels are typically set to zero, or to the mean of the neighbouring values.
    \item The quality of the correlation measurement i.e. the SNR in $G^{(2)}$, depends on the incident photon pair flux on the camera. Remember that the technique we use, described in~\cite{defienne_general_2018}, operates in a regime where we detect more than one pair per frame. This leads to detecting numerous accidental coincidences, which are estimated independently and eliminated by subtraction. If there are too few pairs per image, then each frame is primarily composed of noise. If there are too many pairs, then the number of accidentals is excessively high. Hence, there exists an optimal photon flux. To adjust it, one can vary the exposure time. Even though it can theoretically be predicted from the camera parameters~\cite{reichert_optimizing_2018}, in practice, it is more efficient to find this optimal time by conducting multiple measurements at different exposure times and retaining the best one. In our work, all of the data shown in this work was taken with an exposure time of 2ms.
\end{itemize}

\section{Full description of alignment process}

In this section we describe each step in the alignment process. As shown in Figure \ref{fig:experimentSM}, we use a SLED at 800-820nm as our alignment beam.

\begin{enumerate}

\item \textbf{Co-align SLED and pump.} 
To ensure the SLED will follow the same path as the pairs, it must be aligned with the pump beam. A dichroic mirror (DM) is used to superimpose the pump onto the SLED. Two mirrors in the path of the SLED ($M_2$ and $M_3$ in Fig. \ref{fig:experimentSM}) can fully control the position and angle, whilst one mirror ($M_1$) plus the DM are sufficient for the pump. The two beams can be aligned with each other by ensuring they both pass through two distantly separated irises/pupils. After they are co-aligned, do not move the DM or any mirror before it. Note that if the SLED is at 810 nm, an IR viewing card will be necessary to see the beam. 

\item \textbf{Place mirrors.} 
Now that the SLED is aligned with the pump, place the required mirrors (and the SLM, as this behaves like a mirror also) and align them with the SLED so that the beam follows the desired path. Try to keep the beam as close to the centre of each mirror as possible.

\item \textbf{Place lenses. } Assuming you need to align $n$ lenses, label them $f_1$-$f_n$, where $f_1$ is the first lens after the crystal, and $f_n$ is the last lens, immediately before the camera. We start with $f_n$ and work backwards. The process is as follows: 

\begin{enumerate}
    \item Place the camera in the desired position and ensure the SLED is well-centered on the sensor. Put lens $f_n$ before the camera, and adjust it's position to obtain the tightest possible focus. The centre of the focused and collimated beams should be at approximately the same position. To adjust this, change the lateral position of the lens. With all lenses, ensure they are perpendicular to the optical path. We can swap between imaging and Fourier-imaging the beam by swapping lens $f_n$ with one that has half the focal length, $0.5f_n$. 
    \item Swap lens $f_n$ with lens $0.5f_n$. (Using lenses with threaded mounts is recommended or, if available, swappable magnetic mounts.)
    \item Place lens $f_{n-1}$ and position this to achieve the tightest focus. Again, the beams should be roughly centred at the same position on the sensor.
    \item To place lens $f_{n-2}$, replace $0.5f_n$ with $f_n$ and find the best focus.
    \item For every subsequent lens, first swap $f_n \leftrightarrow 0.5f_n$ and then position the new lens to get the best focus. Always make sure that the collimated beam and focused spots are centred at around the same point at the camera. This ensures that the beam is passing through the centre of each lens, but it is worth double checking this at each step to avoid introducing optical aberrations. 
    \item Finish by placing the lens $f_1$. Generally, this lens has a short focal length, typically between 25 and 50mm. As stated in the Methods, this lens should be mounted on a translation stage for fine control of its alignment. More precise alignment will be done in the following step, but you should still aim for the best alignment possible by hand first. 
\end{enumerate}

\begin{figure*}[ht]
    \centering
    \includegraphics[width=0.7\textwidth]{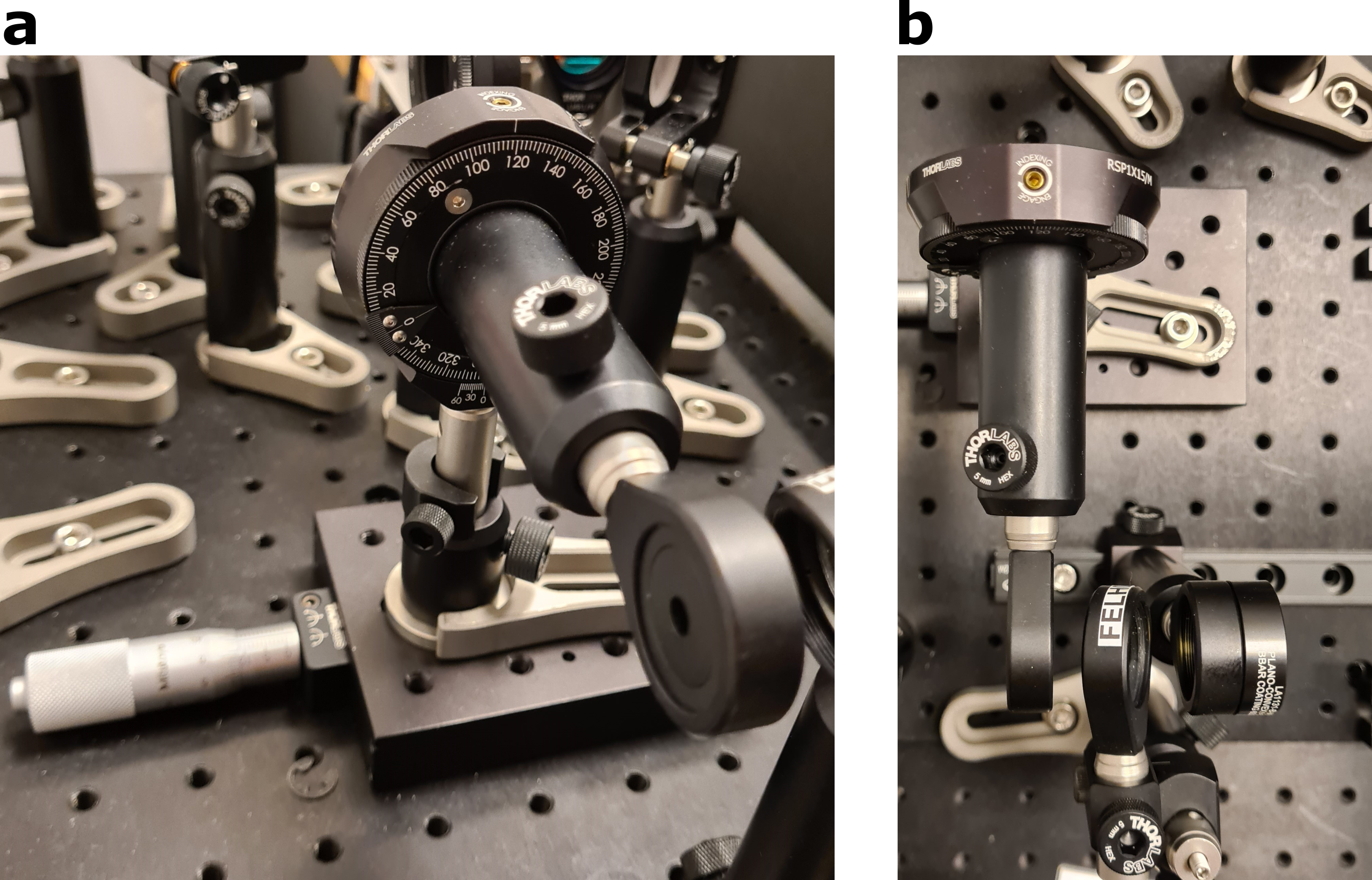}
    \caption{\textbf{Example of a rotatable crystal mount. a,} Image of full mount and translation stage. The crystal is mounted on a circular lens mount and post. This itself is mounted horizontally on a rotation mount to enable simple control over the crystal angle. Finally, everything is mounted on a (manually actuated) translation stage for precise control over the crystal position. \textbf{b,} Top-down image of the same crystal mount to show the crystal followed by a short-pass spectral filter and short focal length lens ($f_1$). The filter is tilted to direct the reflection of the pump to a desired location.  }
    \label{fig:crystalmount}        
\end{figure*}

\item \textbf{Precisely align lens $f_1$ and crystal.} 

The photon-pair correlations are particularly sensitive to the positions of lens $f_1$ and the crystal. The reason is that it is generally necessary to use a lens with a short focal length just after the crystal. Indeed, this allows for a sufficiently large numerical aperture to collect all the k-vectors emitted by the crystal. It is therefore important to align these well.
\begin{enumerate}
 
\item First, choose between the lens with focal length $f_n$ or 0.5$f_n$ in order to image the front focal plane of lens $f_1$ onto the camera. This configuration is called near-field imaging configuration. Now we are imaging the plane in which we want to put the crystal. The crystal is transparent to the SLED, but it will likely have dust/imperfections on its surface. It can be put in roughly the correct plane by getting these imperfections in-focus. If there are no visible imperfections, another option is use a cross target or similar object that can be easily swapped with the crystal without moving the entire crystal mount. For an example of the mount we have used, see Figure \ref{fig:crystalmount}. 

\item Now that  all of the lenses and the crystal are positioned, the SPDC light should be visible on the camera. In the following steps we will measure and use the photon-pairs spatial correlations directly to precisely adjust the positions of the lens and crystal. From here, remove as much background light as possible by covering the setup and turning off all other light sources.
\item Now, put the setup in the far-field imaging configuration i.e. choose between the lens with focal length $f_n$ or 0.5$f_n$ in order to image the back focal plane of lens $f_1$ onto the camera. We will align the lens $f_1$ first since, in the far-field configuration, the correlation width does not depend on the distance between the lens and crystal. 
\item Tilt the crystal around its horizontal axis until a ring (or rings if the crystal is a Type II or paired Type I) is visible on the camera. Slowly tilt the crystal until this ring is collapsed almost to a disk, and fits into an area of at most $200\times200$ pixel. The exact size of this disk will depend on the camera being used but, if it is bigger than $200\times200$ pixels, it is likely too large and will slow down the correlation processing. In this case, it is recommended to reduce the magnification of your imaging system.

\item Do a short $G^{(2)}$ acquisition (i.e. up to few minutes) and compute the sum-coordinate projection. There should be a peak in the centre. If there is a peak, simply find the position of the lens translation stage that maximises it. If there is no peak, move the lens and try again. If, after using the full travel of the stage, you still see no peak, then it is likely a problem with the alignment of other lenses. Check their alignment with the SLED and, if necessary, start again from step 3. 

\item Once the peak of the sum-coordinate projection in the far-field imaging configuration is optimised, move to the near-field imaging configuration. Now, repeat the process above for the position of the crystal instead of the lens, and optimise the peak in the minus-coordinate projection of $G^{(2)}$ instead. 

\end{enumerate}

\item \textbf{Aligning the SLM. }
Ideally, the SLM should be exactly in the Fourier-plane of the camera. However, since it acts in reflection, this is not always easy to accomplish this. After aligning all lenses between the SLM and camera, choose between the lens with focal length $f_n$ or 0.5$f_n$ in such a way as to image a plane close to that of the SLM. Display a high-contrast mask on the SLM with many sharp edges (e.g. a 0/$\pi$ grating). These sharp edges are visible due to diffraction effects, and can be used to position the SLM in the correct plane. Moving the SLM will change the distance between the lenses before and after it, introducing defocus error to the system. Therefore, it is best to use longer lenses directly before and after the SLM so that the depth of focus is long and defocus errors are reduced.

\end{enumerate}

If you have completed all of these steps successfully, your system should be well aligned. The peaks in the near-field and far-field imaging configurations should be similar in size to those in Figures \ref{fig:experiment}d,f.

\section{Full details for SLM calibration}

Here we will give a step by step method to calibrate the pixel response of an SLM. Example Matlab code for this process is given in Ref.~\cite{_see_}. This method requires a camera, a static scattering medium that produces a speckle, and a spatially coherent light source (here we use the SLED). It is assumed that the SLM is functioning and can display arbitrary grayscale images. It is simplest to control the camera and SLM using the same software, e.g. with the same Matlab or Python script. The SLM does not need to be perfectly aligned to perform this calibration step. The calibration is in two steps: the speckle correlation measurement, and the data processing. The measurement goes as follows:

\begin{enumerate}
    \item Align the SLM and camera so that the SLED/laser is incident on the SLM and is visible on the camera. It is convenient to have Fourier-imaging lens between the SLM and camera, but the alignment of this lens does not need to be extremely precise at this stage.
    \item Ensure that the SLM is blank, i.e. displaying a flat phase mask, and place the scattering medium after the SLM. It should be placed so that a wide speckle is shown on the camera. Save an image of this speckle; it will be used as the reference image, denoted $I_0(\vec{r})$. An example speckle is shown in Figure~\ref{fig:speckles}a. 
    \begin{figure*}[ht]
        \centering
        \includegraphics[width=0.8\textwidth]{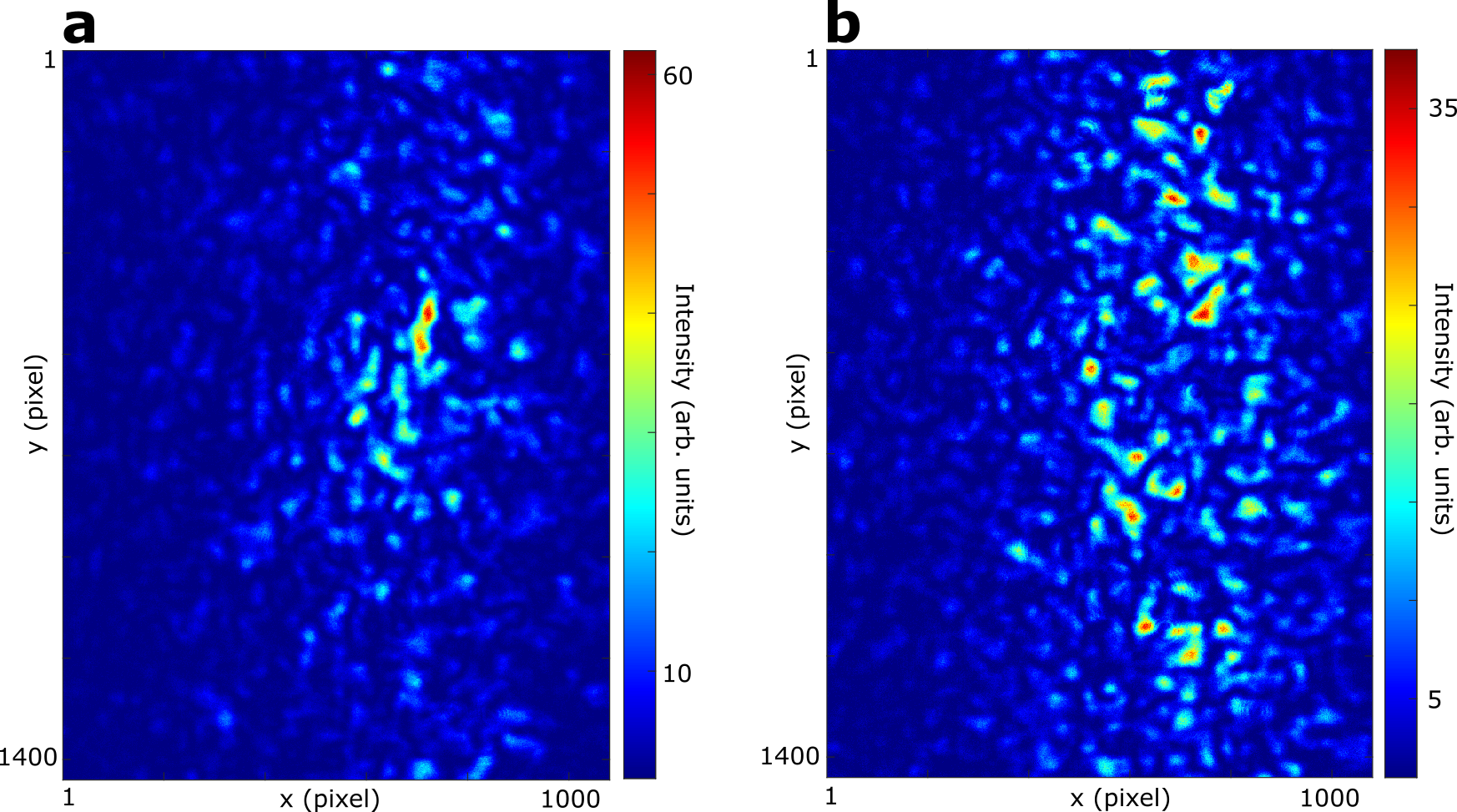}
        \caption{\textbf{ Speckle patterns. a,} Reference speckle pattern measured with a flat phase programmed on the SLM. \textbf{b,} Speckle pattern measured with random subset of SLM pixels set to the gray value $G=117$. Images were acquired with with Thorlabs Zelux 1.6 MP Monochrome CMOS camera. The SLM is an Holoeye Pluto NIR-II.}
        \label{fig:speckles}        
    \end{figure*}
    \item Randomly select approximately half of the pixels in the illuminated region of the SLM. It is best to group the pixels into larger macropixels. This region of the SLM is called the active region. The region containing the pixels that remain at zero is the passive region.
    \item  Now, for each grayscale value (i.e. integer $G$ from 0 to 255), set the selected pixels to this value, display this mask on the SLM, and save an image of the speckle at each step. This image is denoted $I_{G}(\vec{r})$, where $G$ is the grayscale value. An example image of the speckle for non-zero grayscale values is shown in Figure~\ref{fig:speckles}b.
    \item Calculate the correlation coefficient, $M$, between each image ($I_{G}(\vec{r})$) and the reference image ($I_0(\vec{r})$). Here, this is done in Matlab using the function \texttt{corr2}.
    \item Plot the correlation coefficient $M$ as a function of grayscale pixel value, as shown in Figure~\ref{fig:specklecorrplot}a.
    \begin{figure*}[ht]
        \centering
        \includegraphics[width=\textwidth]{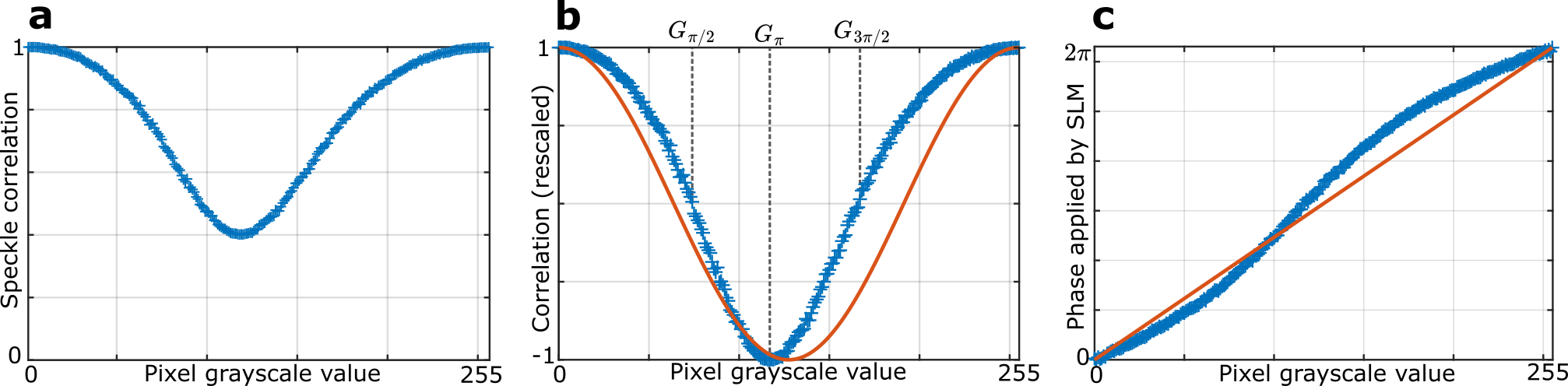}
        \caption{\textbf{Pixel response of SLM measured via speckle correlation. a,} Speckle correlation computed between a reference speckle and speckles from each pixel grayscale value. \textbf{b,} Rescaled speckle correlation vs pixel grayscale value. Grayscale values corresponding to $\pi/2,\pi,3\pi/2$ phase shifts are indicated. Solid red line is the cosine that the data would follow if the pixel response was linear. \textbf{c,} Actual phase shift applied by the SLM, computed from the inverse cosine of the rescaled speckle correlation data. Solid red line is a straight line to illustrate the non-linearity of the pixel response.}
        \label{fig:specklecorrplot}        
    \end{figure*}
\end{enumerate}

As shown in Figure~\ref{fig:specklecorrplot}a, the correlation as a function of the grayscale level $G$ closely resembles that of a cosine. A cosine is what we would expect to observe if the SLM was already perfectly calibrated. More precisely, it should be: 
\begin{equation}
\label{equCalib}
    M=A+B \mbox{cos} \left(\frac{2 \pi}{255} G \right),
\end{equation}
where $A^{final}$ and $B$ are two constants. Indeed, the intensity at a position $\vec{r}$ on the camera results from the interference between a speckle $s_P$ generated by the passive part of the SLM and a speckle $s_A$ generated by the active part. When we phase-shift the active part with respect to the passive part by a global phase $\theta$, then the intensity measured on the camera can be written as follows: $I_\theta(\vec{r}) = |s_P|^2+|s_A|^2 + 2 |s_P s_A| \mbox{cos}\left( \alpha_A(\vec{r})-\alpha_P(\vec{r}) + \theta \right)$, where $\alpha_P(\vec{r})$ and $\alpha_A(\vec{r})$ are the phase components of $s_A(\vec{r})$ and $S_P(\vec{r})$, respectively. Calculating the correlation $M$ between $I_\theta$ and $I_0$ using Matlab's \texttt{corr2} function involves spatially averaging the products $I_\theta(\vec{r})$ and $I_0(\vec{r'})$ for all pairs of positions $\vec{r}$ and $\vec{r'}$. Assuming that each speckle is well developed, then phases $\alpha_P$ and $\alpha_A$ are randomly distributed between $0$ and $2\pi$ across all the camera pixels, leading to the following results:
\begin{align}
    M(\theta) &= \langle I_\theta(\vec{r}) I_\theta(\vec{r'}) \rangle_{\vec{r},\vec{r'}} \nonumber \\
    &= \langle [|s_A(\vec{r})|^2+|s_P(\vec{r})|^2]+[|s_A(\vec{r'})|^2+|s_P(\vec{r'})|^2] \rangle_{\vec{r},\vec{r'}} \nonumber  \\
    &\quad+ 2 \langle  [|s_A(\vec{r})|^2+|s_P(\vec{r})|^2] |s_A(\vec{r'})| |s_P(\vec{r'}) \mbox{cos} \left(\alpha_A(\vec{r'})-\alpha_P(\vec{r'} \right) + \theta) \rangle_{\vec{r},\vec{r'}} \nonumber  \\
    &\quad+ 2 \langle [|s_A(\vec{r'})|^2+|s_P(\vec{r'})|^2] |s_A(\vec{r})| |s_P(\vec{r}) \mbox{cos} \left(\alpha_A(\vec{r})-\alpha_P(\vec{r}) \right) \rangle_{\vec{r},\vec{r'}} \nonumber  \\
    &\quad+ 2 \langle |s_A(\vec{r}) s_P(\vec{r}) s_A(\vec{r'}) s_P(\vec{r'})| \mbox{cos} \left(\alpha_A(\vec{r})-\alpha_P(\vec{r}-\alpha_A(\vec{r'})+\alpha_P(\vec{r'}) \right)  \rangle_{\vec{r},\vec{r'}} \nonumber  \\
     &\quad+ 2 \langle |s_A(\vec{r}) s_P(\vec{r}) s_A(\vec{r'}) s_P(\vec{r'})| \mbox{cos} \left( \theta \right) \rangle_{\vec{r},\vec{r'}} \nonumber 
 \\
     &= A + B \mbox{cos}(\theta).
\end{align}

If the SLM is perfectly calibrated, then $\theta = \frac{255}{2 \pi} G$, leading to equation~\ref{equCalib}. However, in practice this is never the case, and the relationship between $\theta$ and $G$ i.e. $\theta = f(G)$, is not so simple. It is precisely this function $f$ that we are aiming to experimentally measure and determine here. 

To achieve this, we then start form the correlation curve measured in Figure~~\ref{fig:specklecorrplot}a. Firstly, it is necessary to ensure that the SLM implements sufficient phase shifting, i.e.
if the cosine is cut off before reaching a maximum, then the pixels are not modulating all the way to 2$\pi$. Generally, an SLM will come with software to control the voltage that is applied across the pixels. If possible,  use this to adjust the maximum applied voltage so you get one full oscillation, erring towards more than a full oscillation. Once a calibration curve with more than one oscillation is obtained, the curve is unlikely to be a perfect cosine because $f$ is generally not linear. $f$ is determined using the following procedure:
\begin{enumerate}
    \item Rescale the data so that $M$ ranges between -1 and 1. Figure~\ref{fig:specklecorrplot}b shows such a re-scaled curve (blue).
    
    \item Record the gray values $G$ of the first maximum, labelled $G_0$; the minimum, labelled $G_\pi$; and the second maximum, labelled $G_{2\pi}$. Record the $G$ values where $M=0$ (or closest). Label the lower $G_{\pi/2}$ and higher $G_{3\pi/2}$. See Figure~\ref{fig:specklecorrplot}b.

    \item Split the data in half at $G_{\pi}$, so we have $M_{left}=M(G_{left})$ for $G_{left} \in [G_0,G_\pi]$ and $M_{right}=M(G_{right})$ for $G_{right} \in [G_\pi+1,G_{2\pi}]$.
    
    \item Compute $Y_{left}=\mathrm{arccos}(M_{left})$ and $Y_{right}=-\mathrm{arccos}(M_{right})+2\pi$, and plot $Y$ vs $G$, as shown in Figure~\ref{fig:specklecorrplot}c. This is the pixel response of the SLM i.e. the function $f$. 

    \item Fit $Y_{left}$ and $Y_{right}$ as functions of $G_{left}$ and $G_{right}$, respectively. It is usually sufficient to use quadratic polynomials for the models. In the example code, to improve the fit consistency, we shift the right data so that the first point is at the origin. That is, we fit $Y_{right}'=Y_{right}-\pi$ as a function of $G_{right}'=G_{right}-G_\pi$. The results can be re-shifted after the fit. Figures~\ref{fig:fits}ab show examples of fitted models. 

    \item Now you have two models describing the pixel response from 0 to $G_\pi$ and $G_\pi$ to $G_{2\pi}$. If they are quadratics, they are of the form
    \begin{equation}
        Y_{model} = a_1G^2+a_2G+a_3.
    \end{equation}

    For the right-side model, if the fit was done with shifted data, then the coefficients can be redefined: 
    \begin{equation}
        a_1 = a_1, \quad
        a_2 = -2G_\pi a_1 + a_2, \quad
        a_3 = -a_2G_\pi + a_1G_\pi^2 + Y_0
    \end{equation}
    with $Y_0=max(Y_{left})$ ensuring the fits can be merged at $G_\pi$. We use these models to create the function $f$ (e.g. in the form of a Matlab function) that transforms a grayscale level G into its corresponding phase value.

    \item Similarly, we need to create a function corresponding to $f^{-1}$. For this, the models must be inverted. If they are quadratics, then
    \begin{equation}
        G_{model} = \frac{-a_2+\sqrt{a_2^2-4a_1(a_3-Y)}}{2a_1}. 
    \end{equation}
    Now you can define a function that computes $G_{model-left}$ for $Y \in [0,\pi]$, using the coefficients for the left side fit, and $G_{model-right}$ for $Y \in [\pi,2\pi]$ for the right-side fit. See Figure~\ref{fig:fits}c for the final combined model.
    \begin{figure*}[ht]
        \centering
        \includegraphics[width=0.9\textwidth]{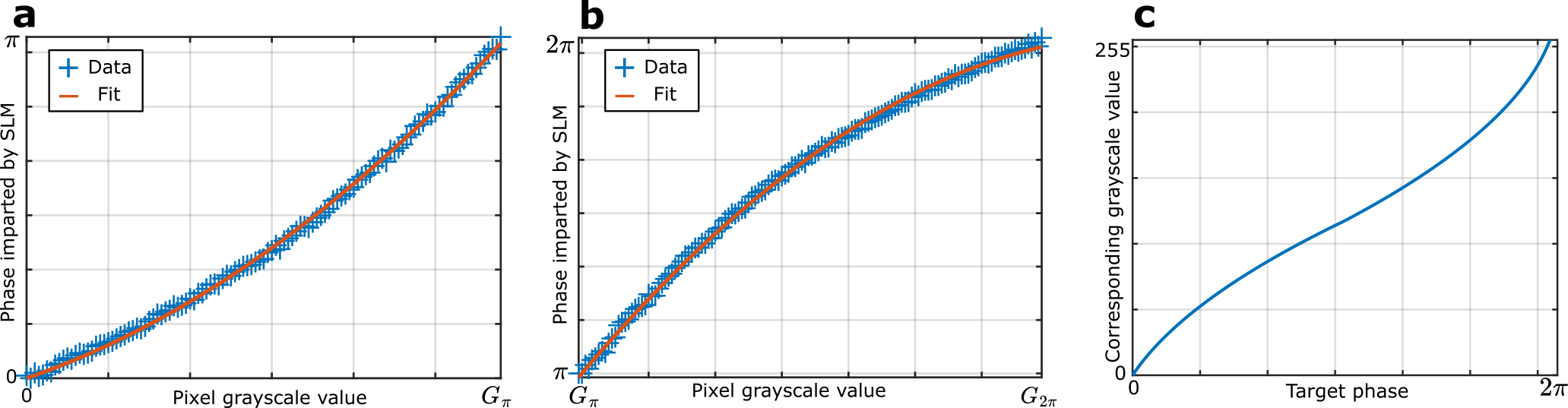}
        \caption{\textbf{Quadratic fits of the measured SLM pixel response. a,b,} Data points and fitted functions for the left and right data, respectively. \textbf{c,} Combined inverse function that maps the target phase to the grayscale pixel value that gives this phase.}
        \label{fig:fits}        
    \end{figure*}
    \item At the end of the calibration, we have two functions corresponding to $f$ and $f^{-1}$ which represent the grayscale values programmed onto the SLM and the actual phase shift that the SLM implements. To check the calibration, repeat the speckle measurements above with grayscale values $G_{pix} = G_{model}(Y)$, with $Y$ being a linear vector of phases from 0 to 2$\pi$. If the calibration was successful, this should result in a perfect cosine shape. 
\end{enumerate}

\end{document}